# Robust Dipolar Layers between Organic Semiconductors and Silver for Energy-Level Alignment

*Tomáš Krajňák,[1] Veronika Stará,[1] Pavel Procházka,[1] Jakub Planer,[1] Tomáš Skála,[2] Matthias Blatnik,[1] Jan Čechal[1,3]\**

[1] CEITEC - Central European Institute of Technology, Brno University of Technology, Purkyňova 123, 612 00 Brno, Czech Republic.

[2] Department of Surface and Plasma Science, Faculty of Mathematics and Physics, Charles University, V Holešovičkách 2, 180 00 Prague 8, Czech Republic.

[3] Institute of Physical Engineering, Brno University of Technology, Technická 2896/2, 616 69 Brno, Czech Republic.

**Abstract**

The interface between a metal electrode and an organic semiconductor (OS) layer has a defining role in the properties of the resulting device. To obtain a desired performance, interlayers are introduced to modify the adhesion and growth of OS and enhance the efficiency of charge transport through the interface. However, the employed interlayers face common challenges, including a lack of electric dipoles to tune the mutual position of energy levels, being too thick for efficient electronic transport, or being prone to intermixing with subsequently deposited OS layers. Here, we show that monolayers of 1,3,5-tris(4-carboxyphenyl)benzene (BTB) with fully deprotonated carboxyl groups on silver substrates form a compact layer resistant to intermixing while capable of mediating energy level alignment and showing a large insensitivity to substrate termination. Employing a combination of surface-sensitive techniques, i.e., low-energy electron microscopy and diffraction, X-ray photoelectron spectroscopy, and scanning tunneling microscopy, we have comprehensively characterized the compact layer and proven its robustness against mixing with the subsequently deposited organic semiconductor layer. DFT calculations show that the robustness arises from a strong interaction of carboxylate groups with the Ag surface, and thus, the BTB in the first layer is energetically favored. Synchrotron radiation photoelectron spectroscopy shows that this layer displays considerable electrical dipoles that can be utilized for work function engineering and electronic alignment of molecular frontier orbitals with respect to the substrate Fermi level. Our work thus provides a widely applicable molecular interlayer and general insights necessary for engineering of charge injection layers for efficient organic electronics.

**Introduction**

Organic electronics is a significant technology for displays and illumination.[1–3] In other fields that utilize organic semiconductors (OSs), e.g., in organic thin-film transistors[4] and organic photovoltaics[5], the large-scale industrial applications are still limited. The performance of fast-switching and high-power organic electronic devices, like OFETs, is often highly influenced by the contact resistance[6–9] originating from the energy level misalignment between a metal electrode and an OS layer.[7,10–12]

Introducing ordered dipolar layers at the metal–OS interface can tune the electrode work function (WF) and the interfacial energy level alignment (ELA) with the OS frontier orbitals (HOMO or LUMO).[13,14] These so-called charge injection layers (CILs) can thus significantly reduce the contact resistance and increase the efficiency of the charge-carrier injection into the OS layer. In this respect, molecular layers exhibiting electric dipoles can act as CILs between metal electrodes and OS layers;[15,16] the dipoles can be either intrinsic to the deposited molecules, formed due to the molecule–substrate charge transfer, or by changing the molecular conformation (e.g., its bending).[13] The self-assembled monolayers (SAMs) were intensively studied in this respect.[13,15–18] The introduction of polar segments into the backbone can provide the desired electric dipoles necessary for WF engineering,[18] but the molecular chains also present a decoupling layer that contributes to the contact resistance between the metal substrate and the OS layer deposited on the top.[17,19–21] In this respect, the OS monolayers demonstrated promising changes of WF with respect to ELA;[10,13,14,22] however, they are prone to interdiffusion or formation of mixed phases with subsequently deposited molecular layers.[10,14,31–33,23–30] A sharp, uniform, and stable interface during the lifetime of the device is required for technological applications of efficient CILs.

Recently, we have shown that monolayers of aromatic carboxylic acids can act as CILs.[34] In this system, the required electric dipoles are localized at the metal–organic interface, which results in removing the tunneling contact between the molecular layer and the metal electrode. However, the employed molecules share the main issue with other molecular species explored for this role: they readily mix with the deposited OS overlayer, which would compromise the performance of potential devices. Here, we show that 1,3,5-tris(4-carboxyphenyl) benzene (BTB, Figure 1a), an aromatic tricarboxylic acid, forms a robust layer that does not mix with deposited OS layers up to temperatures at which OSs either re-evaporate or BTB decompose.

The robust interface can be formed by employing molecules that strongly bind to the surface, like in SAMs. Concerning Ag surfaces, carboxyl terminated SAMs[35–40] show higher structural order than traditionally used thiol-based SAMs.[39,40] Here, a partial charge transfer between molecule and substrate provides a physically robust and electronically strong connection,[9,15,17,41,42] but intermixing with deposited porphyrin and phthalocyanine molecules even below room temperature was reported.[43] In addition, the strong OS molecule–metal interaction induces undesirable changes to surface and OS film microstructure and substantial modification of interfacial electronic structure, which can profoundly impact contact and channel resistance and overall device performance.[7,19,41] Some strongly interacting small organic molecules, like F4-TCNQ and F6-TCNNQ, may form an organometallic layer with silver with a thickness up to 50 nm, which is stable with respect to subsequent deposition of pentacene layers[44] but is still far from an ideal case.

While providing favorable properties with respect to ELA, planar weakly adsorbing OS molecules are more prone to intermixing with subsequently deposited molecular layers. One of the possibilities is to change molecular functional groups or their number to strengthen organic–metal interaction and, thus, the first layer stability. In this respect, changing the molecular structure of pentacene oxo-derivatives from 6,13-pentacenequinone (P2O, featuring two oxygens) and 5,7,12,14-pentacenetetrone (P4O, 4 oxygens) leads to the change of adsorption behavior on Ag(111) from physisorption of P2O to chemisorption of P4O.[31] In this case, the P4O layers were resistant to intermixing with subsequently deposited copper phthalocyanine (CuPc). The other possibility to obtain a semi-stable bilayer is to use 3,4,9,10-perylene-tetracarboxylic-dianhydride (PTCDA), which is stable against the mixing with subsequently deposited CuPc[45] or tin phthalocyanine (SnPc)[25]. In these cases, a kinetic barrier exists regarding interlayer exchange in both CuPc/PTCDA/Ag and PTCDA/CuPc/Ag stacking orders, with a primary parameter governing stability at lower temperatures being the adsorption energy per area of the individual molecules.[30] However, beyond the onset of desorption, the decisive parameter becomes the adsorption energy per molecule, and the preferred occupancy of the first layer can change.

Our previous study introduced aromatic carboxylic acids as dipolar layers.[34] We have shown that the employed 4,4'-biphenyl dicarboxylic acid (BDA, Figure 1b) molecule can gradually deprotonate in direct contact with silver surfaces either thermally[46–49] or by low-energy electrons[50], thus providing a possibility to finely tune the ELA. While considerable shifts in the WF and energy levels of deposited molecules up to 0.8 eV were induced, our later experiments have shown that it is prone to mix with pentacene layers deposited on top. In the present paper, we show that extending the molecule to three carboxylic groups results in a robust monolayer that does not mix with subsequently deposited OS molecules, i.e., pentacene (Figure 1c), a prototypical high mobility OS,[51] HM-TP (Figure 1d), and HAT-CN (Figure 1e), an electron donor and acceptor, respectively. Our DFT calculations show that the robustness is of a thermodynamic origin: the compact layer presents the lowest energy state. Thus, the molecular monolayers of fully deprotonated BTB form a viable platform on the path toward the ohmic contacts between electrodes on OS layers.

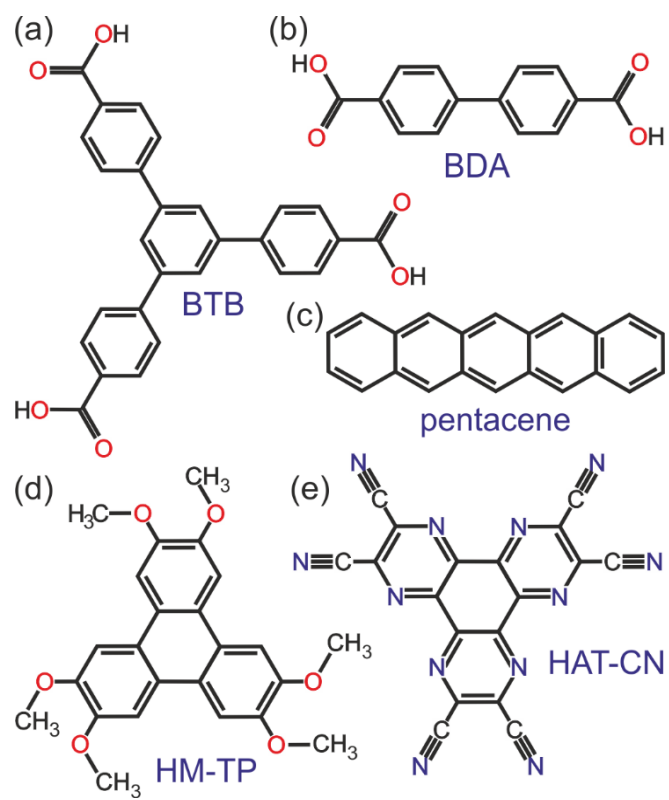

**Figure 1:** Chemical structure of organic molecules explored in this work. (a) 1,3,5-tris(4-carboxyphenyl) benzene (BTB); (b) 4,4'-biphenyl dicarboxylic acid (BDA); (c) pentacene; (d) hexamethoxy-triphenylene (HM-TP); and (e) hexaazatriphenylene-hexacarbonitrile (HAT-CN).

**Results and Discussion**

We have performed experiments for two low-energy facets of the silver surface: Ag(111) and Ag(100). As the results are similar on both surfaces, we will focus our description on Ag(111) and give the results for the other facet in the Supporting Information. In the following, we will first show synchrotron radiation photoelectron spectroscopy results for gradual deprotonation of BTB and show that with respect to WF changes and ELA, the BTB behaves consistently with our earlier results on BDA.[34] Then, we will discuss the obtained STM and LEEM data for submonolayer and full monolayer coverages of the fully deprotonated molecule (marked as δ-BTB in the following), demonstrating that, contrary to BDA, the compact monolayer of the fully deprotonated BTB molecules covers the whole substrate surface (further referred to as compact δ-BTB layer) and is easily achievable. The compact δ-BTB layer is stable against mixing with pentacene, HAT-CN, and HM-TP, typical examples of organic semiconductors: we will show a thermodynamic preference for the formation of pentacene–BTB mixed phases for submonolayer coverages and demonstrate the robustness of the compact δ-BTB layer against structural and chemical changes. Our DFT calculations reveal that the compact δ-BTB layer possesses the lowest energy with respect to other possibilities, so they are robust from the thermodynamic point of view under the UHV conditions.

**Photoelectron Spectroscopy**

Aromatic carboxylic acids deprotonate (i.e., lose hydrogen from carboxylic –COOH groups) upon contact with metal substrates (except for gold) under UHV conditions.[52] This chemical reaction occurs below room temperature for most metals, including Cu[53]. The reaction is kinetically restricted on Ag surfaces, and annealing at elevated temperatures (30–50 °C) is usually required to obtain partially deprotonated molecular phases within minutes.[49] However, significantly higher temperatures (~200 °C) are necessary to achieve complete deprotonation because stable molecular phases hinder the deprotonation reaction.[49] We have followed the deprotonation of BTB on both Ag(111) and Ag(100) substrates by photoelectron spectroscopy employing synchrotron radiation.

The O 1s spectrum of 1 monolayer (ML) of as-deposited BTB molecules on Ag(111) shown in Figure 2a can be fitted by two pairs of peaks (light blue and blue; light green and green). As detailed in Supporting Information, Section 1, we assign these peak components to carboxyl groups in two distinct binding motives. The intensity ratio of these pairs is 2:1. The higher binding energy component from each pair (highlighted by a lighter color in Figure 2a) is associated with hydroxyl oxygen (C–$\underline{O}$H) and the darker one with carbonyl oxygen (–C=$\underline{O}$) of the carboxyl group (–COOH) by comparison with previous works.[46,48,49] Two distinct pairs of peaks point to the existence of two different chemical environments of the carboxyl groups; these are probably associated with the ribbon-like structure of the compressed as-deposited phase (see Figure S3 in Supporting Information, Section 2).

During the annealing at progressively higher temperatures, a new component associated with carboxylate groups[46,48,49] appears in the spectra and grows in intensity (red component in Figure 2a). The relative intensity of this peak is a measure of the degree of deprotonation of carboxylic groups (i.e., the fraction of deprotonated carboxyl groups with respect to all carboxyl groups) in the BTB layer. Figure 2b shows the evolution of the degree of deprotonation with annealing temperature for both Ag surfaces. On both surfaces, BTB molecules gradually deprotonate; on Ag(100), the deprotonation occurs at lower temperatures (consistently with BDA[49]), and complete deprotonation is observed at 170 °C whereas on Ag(111), it is reached at 240 °C. For the Ag(111) substrate, this temperature is already very close to the threshold for the decarboxylation of BTB molecules, i.e., a complete removal of carboxyl groups that occurs around 250 °C for both surfaces. Above this threshold, the XPS data show a decrease of oxygen-related signal while the C 1s peak associated with phenyl rings keeps its intensity and shifts back to higher binding energies – 284.7 eV at Ag(111) and 284.9 eV at Ag(100) – as the carboxylate-related dipoles cease to exist. Disordered polymer-like networks remain on the surface, as observed by STM (Figure S4 in Supporting Information, Section 2). We observe (Figure 2b) that the fully deprotonated δ-BTB phase is stable in a broad window of temperatures of 170 – 250 °C on Ag(100) but only in a relatively narrow range of 235 – 250 °C on Ag(111).

The WF measured after each annealing is displayed in Figure 2c. The WF was determined from the position of the secondary electron cut-off.[34] Due to the push-back effect, with increasing BTB coverage, the WF decreases below 4.1 eV on both surfaces.[14,34] At higher temperatures, the gradual deprotonation leads to the formation of interfacial dipoles, and the WF increases again,[34] reaching 4.61 eV on Ag(111) and 4.49 eV on Ag(100). A different WF of pristine surfaces explains this difference: the measured values were 4.38 and 4.48 eV for Ag(100) and Ag(111), respectively; their values are within the uncertainty interval of reported values, i.e., (4.36 ± 0.05) eV for Ag(100) and (4.53 ± 0.05) eV for Ag(111).[54]

To give a deeper insight into the adsorption-induced WF change, we characterized the structural and electronic properties of an δ-BTB/Ag(111) interface with ab-initio calculations following the procedure described elsewhere.[34] The change in the WF is attributed to the sum of the surface dipoles across the reorganized Ag substrate and the δ-BTB layer and the redistribution of the charge density at the interface resulting from molecule-substrate interaction. The smallest contribution of –0.06 D per BTB molecule arises from the substrate rearrangement. As shown in Figure 2d, subtle changes in the topmost silver layer give rise to this contribution. The intramolecular dipole moment caused by a bending of the molecule and shift of negatively charged oxygen atoms towards the substrate is calculated to be –2.89 D. Finally, the interface dipole moment calculated from plane-averaged charge density difference contributes with +3.57 D per BTB molecule. This contribution arises from a charge density difference plotted in Figure 2d, which shows electron depletion from the topmost silver layer and accumulation in the oxygen layer situated 2.2 Å from the substrate. The overall surface dipole density of the δ-BTB layer thus results in 0.62 D per BTB molecule, causing a 0.14 eV increase in WF from 4.49 eV for the pristine Ag(111) surface to 4.63 eV for the δ-BTB layer of Ag(111) surface in a perfect alignment with experimental observations.

In addition, we have measured the shift of energy levels for as-deposited (α-BTB) and fully deprotonated (δ-BTB) layers by analyzing the positions of phenyl-ring-related components of the C 1s peak for the first and second molecular layer; the procedure is described in our previous work.[34] In Figure 2e, we have plotted the position of C 1s peak within the first layer for BTB together with values obtained for several BDA molecular phases obtained previously.[34] . The BTB data fit the previously established linear trend between the measured WF and core-level positions. The position of core levels experiences the same shift as frontier orbitals in the case of vacuum level alignment.[31,55]

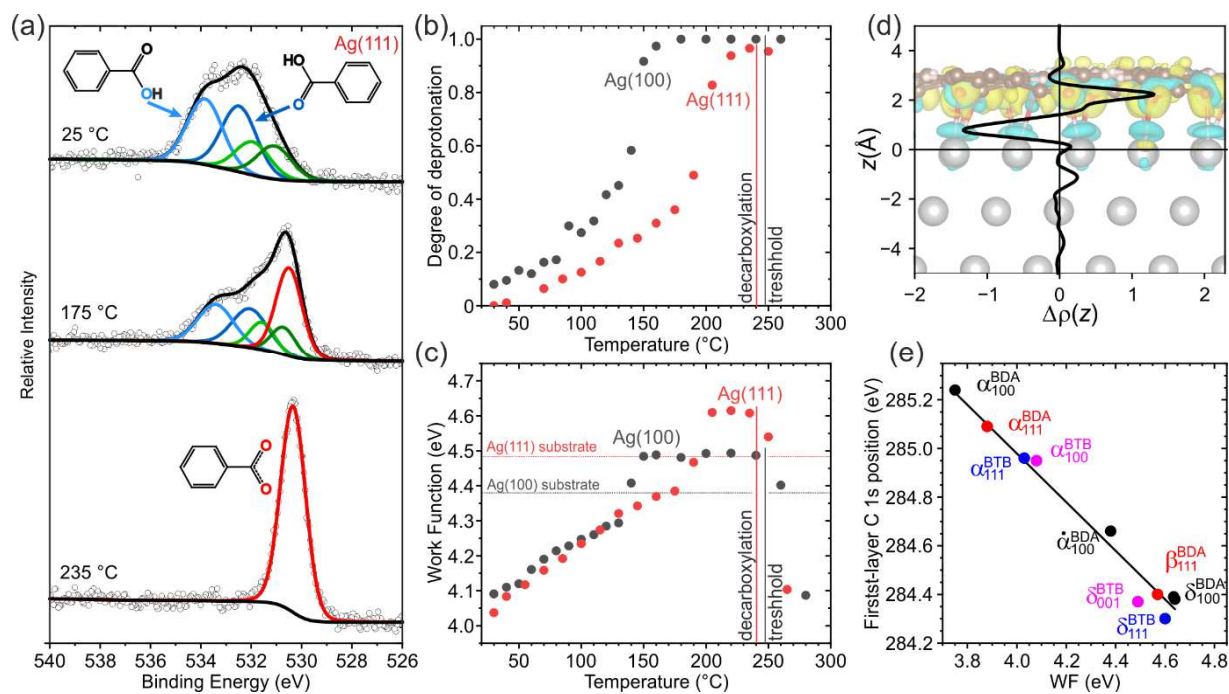

**Figure 2:** Changes in electronic properties of BTB/Ag(111) during its gradual deprotonation. (a) Examples of O 1s spectra recorded on the as deposited phase at 25 °C, after annealing at 175 °C and annealing at 235 °C. (b) Degree of deprotonation of BTB carboxylic groups as a function of annealing temperature for both Ag(111) and Ag(100) surfaces. The vertical lines mark the threshold for decarboxylation, beyond which the decrease of O 1s peak intensity and decrease in sample WF is observed. (c) Sample WF as a function of annealing temperature for both Ag(111) and Ag(100) surfaces. The vertical lines mark the decarboxylation threshold; the horizontal lines mark the measured WF of the bare substrate surface. (d) Plane-averaged difference in charge density along the z-direction perpendicular to the δ-BTB/Ag(111) interface. The relaxed structure and the 3D isosurface of the charge density difference are depicted in the background. Silver, carbon, oxygen, and hydrogen atoms are in gray, brown, red, and white, respectively; electron depletion is colored blue, and accumulation yellow. (e) Position of the C 1s peak associated with phenyl rings within the first BTB molecular layer plotted as a function of the sample WF compared with earlier results for BDA.[34] The lines have a slope of -1, whereas the fitted experimental values have a slope of 1.03 ± 0.06.

**STM and LEEM investigation of δ-BTB layers**

STM and LEEM experiments have been carried out in our home UHV cluster system. We have explored submonolayer and full monolayer coverages of the fully deprotonated BTB phase (δ-BTB) on both Ag(111) and Ag(100) surfaces. As the results are very similar for both substrates, we will present only data for Ag(111) in the main text, and the data for the Ag(100) surface are given in Supporting Information, Section 3.

To obtain the compact δ-BTB layer, the as-deposited BTB layers were annealed at temperatures necessary for the full deprotonation given in the previous section; the full deprotonation was proven by in-situ XPS via the presence of a single O 1s peak component at 530.5 eV (Figure S9, Supporting Information, Section 4), which is consistent with the synchrotron radiation data presented above. The structural evolution of molecular phases during gradual deprotonation was already described earlier in an STM work by Ruben et al.[56] Our data of the as-deposited as well as partially deprotonated molecules are generally in line with their observations. In addition, we could reveal a high degree of complexity in the phase transformations in which the coverage and deposition rate play a significant role. However, a more detailed description of this is beyond the scope of this work.

The molecular-scale topography of the δ-BTB phase obtained by STM shows the BTB molecules as bright protrusions of three-point stars in a hexagonally close-packed structure. The carboxylate (–COO) groups situated at the tips of the stars thereby point to the centers of neighboring molecules. This is shown in detail in Figure 3a, with the superstructure unit cell highlighted as a black rhombus. This phase was originally denoted as phase III with a degree of deprotonation of 2/3.[56] However, our combined STM, XPS, and LEEM data clearly indicate that this phase is fully deprotonated. Figures 3b–d show image details of the molecular structure on step edges and domain boundaries. Figure 3b shows the arrangement of the molecules along and over a single substrate step edge. All the molecules at the upper side of the step edge show the same structure, with one point of the star protrusion missing. The arms of BTB molecules are partially flexible and thus can bend towards the lower terrace. This behavior is even more pronounced at a kink site where the BTB seems to have lost a complete arm. The kink also exactly follows the BTB shape and thus allows seamless growth of the compact δ-BTB layer over the step edge. In this way, the single domain extends over several mono-atomic steps, as shown in Figure 3d. This is evident from a line scan (see inset of Figure 3d) along the white line that shows a step height of ~244 pm, which is slightly higher but in line with the step height of the Ag(111) substrate (236 pm). The molecular arrangement at the domain boundary is shown in Figure 3c. In our STM images, we have seen 4 orientations of molecules. In particular, we identify two different domain orientations (see Figure 3c, regions I and III) and two structural domains (I and II) that share the same unit cell but consist of molecules with orientation mirrored along the unit cell's main diagonal. The calculated DFT model shown in Figure 3e is fully consistent with our STM data. It provides a deeper insight into the interface structure. BTB molecules are rotated by 10.5° with respect to the high-symmetry direction of Ag(111) substrate. The most common site for oxygen atoms to adsorb is in the on-top position, while one of the six oxygen atoms is situated in the bridge position (marked with a blue arrow).

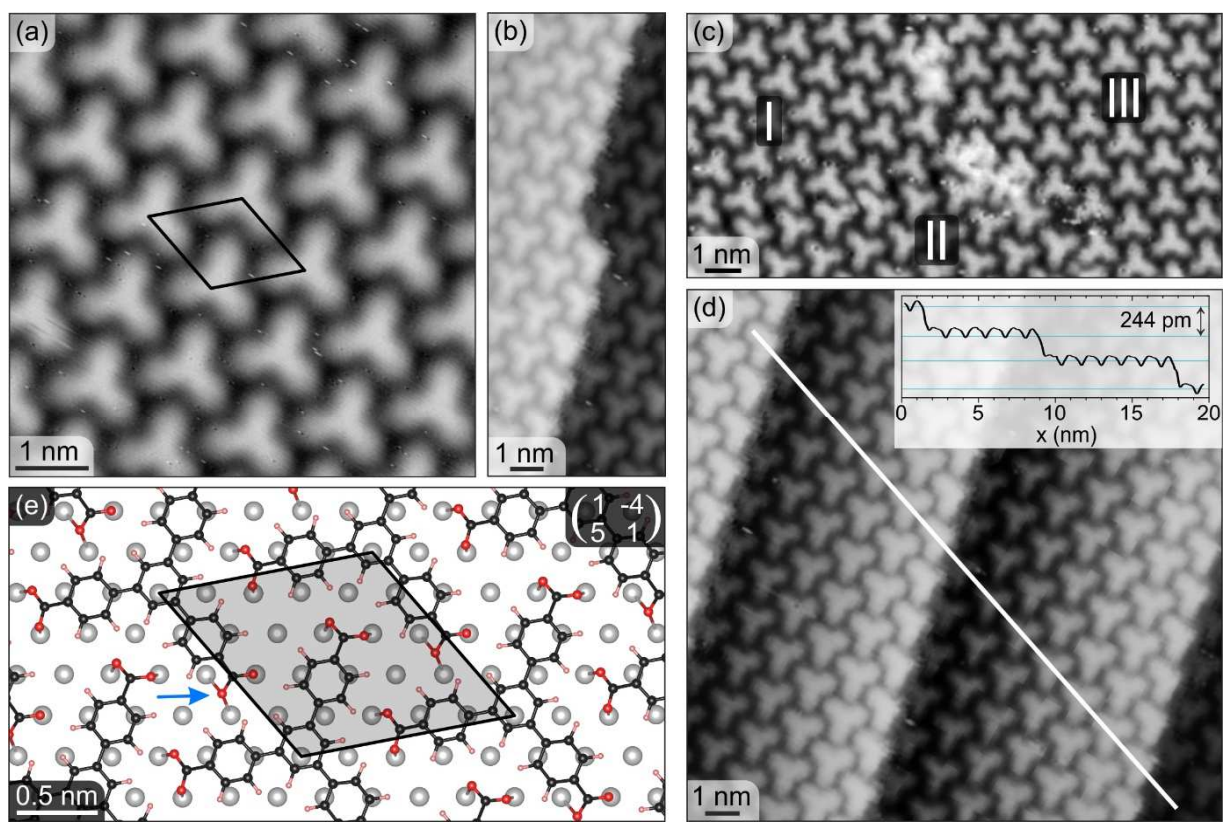

**Figure 3:** δ-BTB phase on Ag(111) surface. (a–d) Detailed STM images of the δ-BTB phase: (a) on a flat terrace showing the structure of the phase with the unit cell is highlighted as a black rhombus; (b) growth of δ-BTB molecules across one step-edge and an extended kink; (c) boundary of three δ-BTB domains marked I, II and III (I and III are different rotational domains whereas in I and II show a mirror symmetry); and (d) the δ-BTB phase with a single-orientation extending over several terraces; the inset shows a line scan along the white line indicated. Scanning parameters for all STM images: 1.4 V, 50 pA. The full-size images are given in Supporting Information, Section 5. (e) DFT-based model of the δ-BTB phase showing the molecular arrangement on the Ag(111) surface: C: black, O: red, H: light red, Ag: gray. The highlighted unit cell is positioned in the same way as in (a); it features one molecule per unit cell and shows the adsorption positions of the three terminal carboxylate groups. Two carboxylates are aligned such that both O atoms adsorb in an on-top position. In the third carboxylate group, only one of the O atoms is in an on-top position, whereas the second is in a bridge position (highlighted by a blue arrow).

The LEEM measurements shown in Figure 4 provide real and reciprocal space views on sample morphology and structure at the mesoscale. The large area diffraction pattern of the δ-BTB phase is presented in Figure 4a. The microdiffraction measurement reveals that the δ-BTB phase exists in two rotational domains on the Ag(111) surface: the model of large-area diffraction pattern decomposed into two single-domain diffraction patterns is given in Figure 4b. The modeling of the δ-BTB diffraction pattern provides a $\begin{pmatrix} 1 & -4 \\ 5 & 1 \end{pmatrix}$ unit cell (in this work, all the superstructure unit cells are given in the matrix notation). These two domain orientations were also identified in our STM images; see Figure 3c. In addition, each of these domains has an additional structural domain with the same unit cell but a mirrored orientation of molecules within them (see, e.g., Figure 3c). In the microdiffraction data and diffraction model, these two mirrored domains are indistinguishable.

The bright-field LEEM image (Figure 4c) portrays submonolayer coverage δ-BTB islands as a bright area on the dark background, which represents the bare substrate; the average area of the BTB islands is $0.3 \pm 0.1$ μm$^2$. LEEM dark-field imaging, in which the image is formed only by electrons associated with a single diffraction spot different from the (0,0), allows real-space visualization of the rotational domains. For submonolayer coverage, individual δ-BTB islands grow in single domain orientation. However, if the surface is completely covered (Figure 4d), we observe a larger number of smaller rotational domains within the δ-BTB layer; the upper bound of the average area of these domains is $0.011\pm0.004$ μm$^2$, i.e., much smaller compared with the island size in the submonolayer coverage. The smaller domain size is probably caused by a limited BTB transport via surface diffusion, which is hindered in the full monolayer.[48] Still, the δ-BTB surface shows a superior long-range order with a minimum of defects as the two domains are well matched at their boundary (see Figure 3c), and single domains extend across the step edges (see Figure 3b and d). On the other substrate facet, Ag(100), the structure of the compact δ-BTB layer is very similar to Ag(111) presented above: the molecular packing is the same with three BTB molecules per unit cell commensurate with the substrate and the area per molecule differs by 2 % (see Supplementary Information, Section 6, for details).

We have tested the applicability of the compact δ-BTB layer as a CIL for OSs. In the following, we will describe the experiments with pentacene; the experiments with HAT-CN and HM-TP (Figure 1c-e) are given in Supporting Information, Section 7.

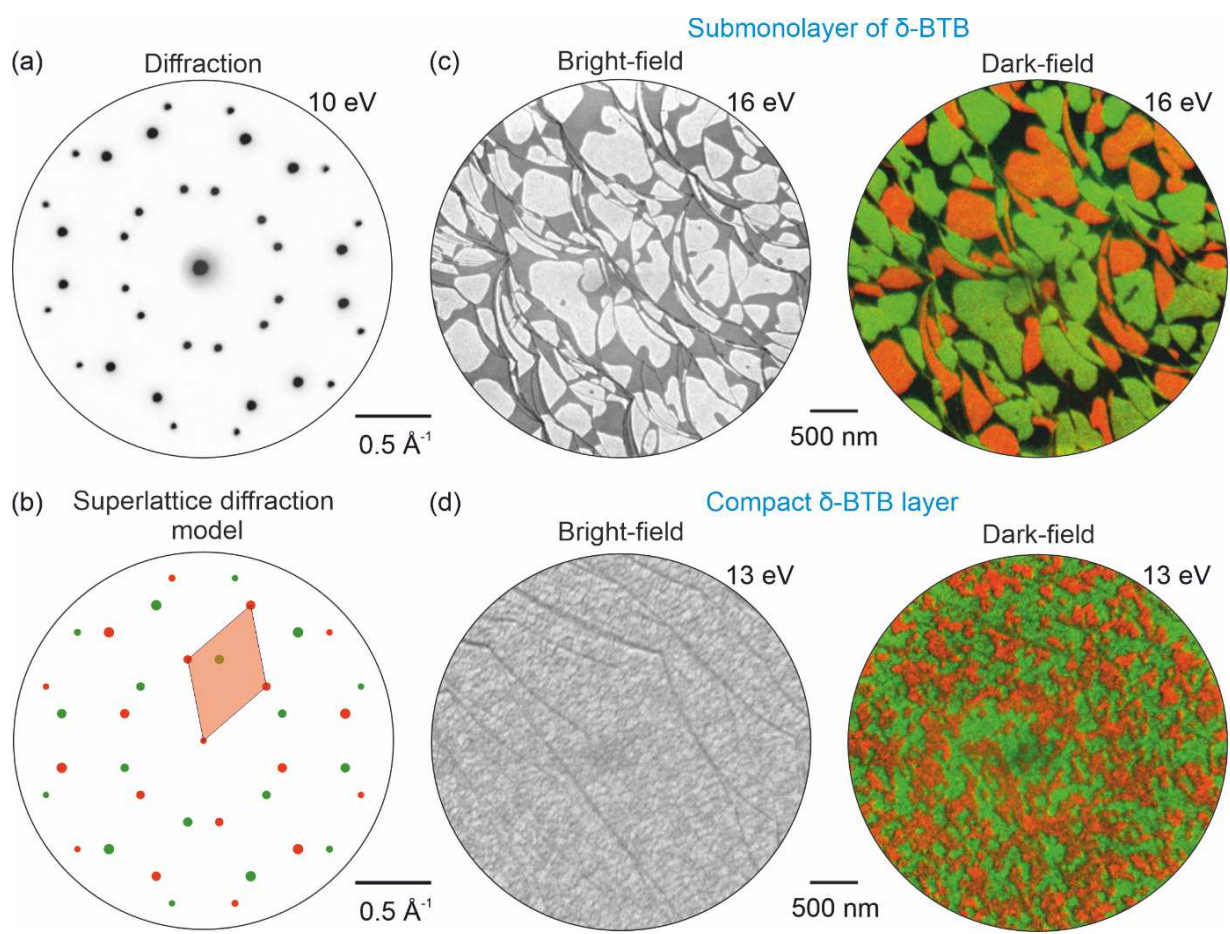

**Figure 4:** LEEM analysis of the δ-BTB phase on the Ag(111) surface. (a) Large-area diffraction pattern taken at 10 eV primary electron energy. (b) Superlattice diffraction model of the δ-BTB layer showing the composition from two single-domain diffraction patterns. (c) Bright and dark-field images taken at the submonolayer BTB coverage showing δ-BTB islands; the green and red color in the dark-field image is associated with a particular rotational domain given by the microdiffraction model in panel (b). (d) Bright and dark-field images of the compact δ-BTB layer; the color-coding is the same as in (c).

**Formation of mixed pentacene–BTB phases at submonolayer BTB coverage**

At 1 ML coverage, δ-BTB molecules form a compact layer, which is stable against mixing with subsequently deposited organic semiconductor molecules. However, this changes in the submonolayer regime, where pentacene forms mixed phases with BTB. Deposition of 0.5 ML of pentacene and 0.5 ML BTB molecules on Ag(111) substrate and subsequent annealing (170 °C, 30 minutes) results in the formation of mixed pentacene–BTB phases. During the annealing, the BTB molecules deprotonate, and the pentacene–BTB mixed phases appear upon cooling. The bright-field image in Figure 5a shows molecular islands of the mixed phase (brighter areas) covering approximately 1/3 of the substrate, whose size and shape are restricted by the substrate step edges. The remaining molecules are present in molecular gas or disordered phases. The diffraction pattern (Figure 5b) measured on these islands is distinct from those observed for pure BTB phases. Employing ProLEED Studio to model the diffraction pattern (Figure 5c), we find the associated unit cell as $\begin{pmatrix} 3 & -13 \\ 15 & 8 \end{pmatrix}$. The STM analysis reveals that this phase comprises two close-laying pentacene molecules sandwiched between two δ-BTB molecules, as visualized in Figures 5d–f, giving the 1:1 ratio of pentacene and BTB. Moreover, the pair of pentacene molecules is tilted at the corners of the unit cell with respect to the two pentacene pairs in the interior, as shown in Figures 5d, e.

We note that the resulting molecular arrangement in mixed phases can be affected by the initial ratio of deposited molecules. In another experiment, we deposited 0.8 ML of pentacene BTB and 0.5 ML of BTB molecules and annealed the sample at 170 °C. After cooling, a wheel-like mixed phase with a 2:1 ratio was formed; see details in Supporting Information, Section 8.

Mixed pentacene–BTB phases were formed in all experiments with a submonolayer coverage of BTB molecules. Mixed phases can be formed in several ways. One way is to deposit both molecules on the surface and obtain the mixture with subsequent annealing. Another possibility is to first create δ-BTB, deposit pentacene, and anneal the system afterwards. The main parameters influencing the resulting structure for both procedures are the concentrations of both types of molecules on the surface and the annealing temperature, which needs to be high enough to reach the full deprotonation of the BTB molecules or dissolve δ-BTB islands but still below the decarboxylation and desorption onset.

These experiments with submonolayer BTB coverage indicate a thermodynamic preference for forming mixed molecular phases from pentacene and BTB over the separate pure molecular phases.

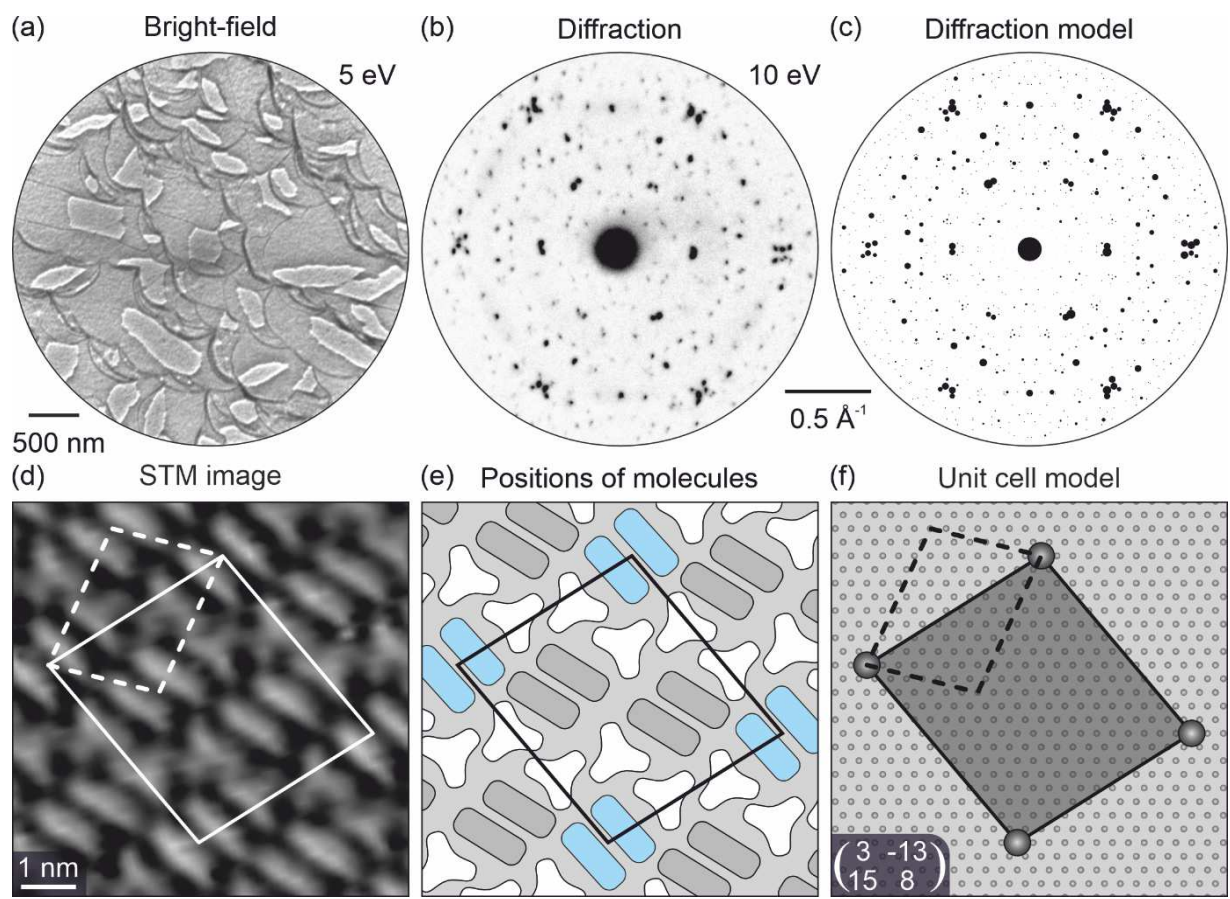

**Figure 5:** Pentacene-BTB mixed phase on Ag(111). (a) Bright-field image of the mixed phase formed by deposition of 0.5 ML pentacene and 0.5 ML BTB molecules and subsequent annealing at 170 °C. (b) The diffraction pattern originating from the mixed phase is shown in (a). (c) The diffraction model of the mixed phase shown in (b). (d) STM image of mixed pentacene-BTB phase with highlighted unit cell (solid line) and an apparent unit cell used for DFT calculations (dashed). (e) Schematics of arrangement of molecules within the unit cell obtained from STM. (f) Position of superstructure unit cell with respect to Ag(111) substrate.

**Pentacene deposition on the compact δ-BTB layer**

We have deposited pentacene on a sample covered by a compact δ-BTB layer. After the pentacene deposition, the LEEM bright-field image shows a compact δ-BTB layer covered with pentacene islands (Figure 6a) that appear as darker areas on a bright δ-BTB background. A LEEM dark-field analysis of δ-BTB spots given in Figure 6b reveals that BTB molecules still cover the whole surface, and the pentacene overlayer attenuates the δ-BTB signal. Figure 6c shows a diffraction pattern that is a superposition of a pronounced diffraction pattern associated with a crystalline overlayer, likely associated with pentacene, and a faint pattern associated with the δ-BTB layer located below.

Annealing the sample at 100 °C for 15 min induces the complete desorption of pentacene: the LEEM/LEED results (Figure 6d–f) show a compact δ-BTB layer similar to that before the pentacene deposition. We did not reveal any sign of the formation of mixed phases comprising BTB and pentacene. XPS spectra of C 1s and O 1s taken before (red) and after (blue) pentacene deposition and sample annealing (green) are given in Figure 7. After pentacene deposition, we observe an increase in the intensity of the C 1s peak, which decreases to the original one after annealing. The O 1s peak shows only a slight change both after deposition and annealing, as pentacene comprises only carbon atoms.

Based on XPS and LEEM observations, we conclude that the full δ-BTB layer is robust against the mixing with pentacene. This robustness can be either of thermodynamic or kinetic origin. The fact that pentacene and BTB form mixed phases suggests that forming bonds between pentacene and δ-BTB molecules is favorable, which indicates the kinetic origin of the robustness. However, the DFT analysis given below shows the opposite, as at coverages approaching a full monolayer, the adsorption energy per unit area dictates the thermodynamic stability of the compact δ-BTB layer.

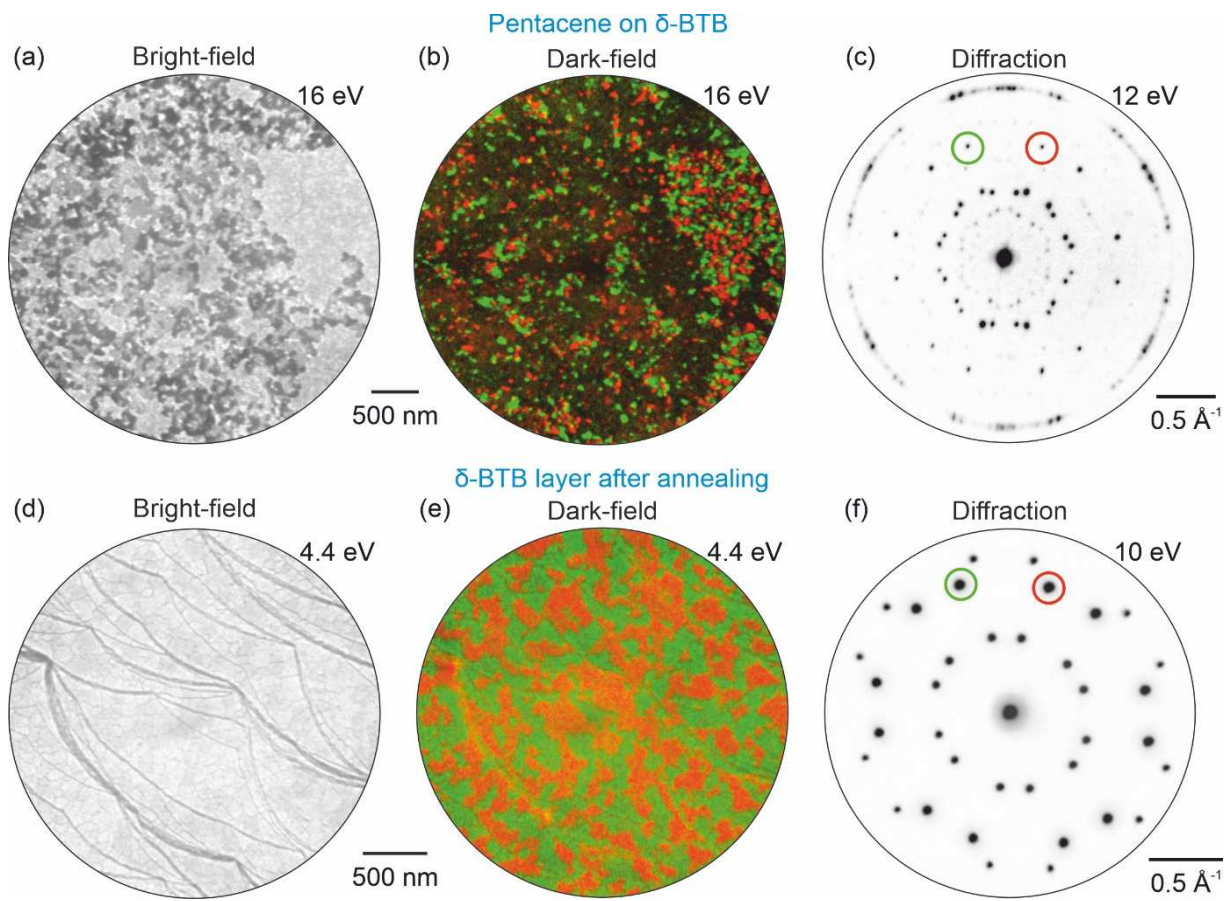

**Figure 6:** LEEM analysis of pentacene on compact δ-BTB layer on Ag(111). (a) LEEM bright-field image showing δ-BTB domains (brighter areas) partially covered by pentacene (darker areas). (b) Composition of dark-field images measured for the two δ-BTB orientational domains; the employed diffraction spots are marked in (c). Only areas without overlayer show a considerable intensity from the δ-BTB layer spots. (c) Diffraction pattern measured on pentacene deposited on the compact δ-BTB layer showing the sum of a faint pattern associated with δ-BTB and the one associated with the overlayer. (d, e) Bright and dark-field images obtained after annealing show a compact δ-BTB layer present on the surface. (f) Diffraction pattern measured after annealing showing a bright δ-BTB pattern without any additional spots.

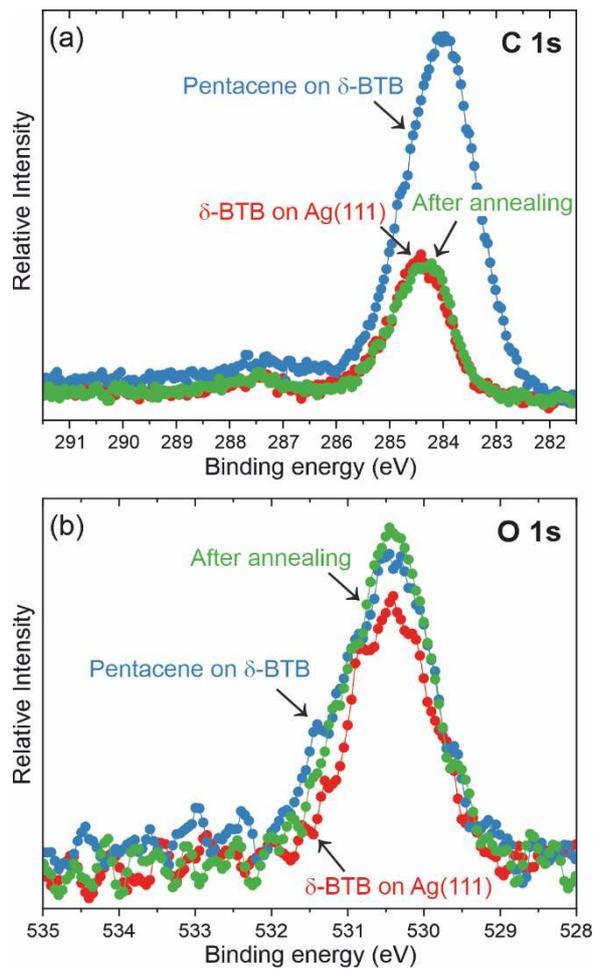

**Figure 7:** XPS analysis of pentacene on the compact δ-BTB layer on Ag(111). (a) C 1s and (b) O 1s spectra measured on the compact δ-BTB layer (red), after pentacene deposition (blue), and subsequent sample annealing at 100 °C (green).

**DFT calculations: Thermodynamic stability of the intermixed phase and δ-BTB layer**

In the following, we demonstrate the energetic preference of the mixed pentacene-BTB phase in the submonolayer coverage and the preference for the δ-BTB phase at full monolayer coverage. In both cases, the decisive factor that determines the stability is the adsorption energy of a molecule per unit area calculated as

$$\gamma = \frac{E_{\text{mol+sub}} - (E_{\text{mol}} + E_{\text{sub}})}{S}, \quad (1)$$

where $E_{\text{mol+sub}}$ is the total energy of a molecular phase on a substrate with area $S$, $E_{\text{mol}}$ denotes gas-phase energies of δ-BTB and pentacene molecules and $E_{\text{sub}}$ is the total energy of a bare substrate (see Supporting Information, Section 9, for the results if a protonated BTB in the gas phase is used as an energy reference). Monolayers of pentacene and δ-BTB were modeled with periodic boundary conditions using the Ag(111) supercells given by $\begin{pmatrix} 6 & 0 \\ 2 & -3 \end{pmatrix}$ and $\begin{pmatrix} 1 & -4 \\ 5 & 1 \end{pmatrix}$, respectively. Due to its size, the real superstructure unit cell for the pentacene-BTB mixed phase $\begin{pmatrix} 3 & -13 \\ 15 & 8 \end{pmatrix}$ is approximated by a smaller, apparent unit cell of $\begin{pmatrix} 6 & -2 \\ 9 & 10 \end{pmatrix}$ depicted in Figure 5f and in Figure S18c. This induces ~3% strain in the shorter surface vector and 2.5 % angular strain. Reference energies for the silver substrate were calculated for each supercell separately. The resulting stabilities, i.e., absolute adsorption energies and energies per unit area, for pentacene, δ-BTB, and the mixed phase on Ag(111) substrate are summarized in Table 1. We note that, in line with experiments, our DFT calculations do not show any surface reorganization, which is not favored due to a relatively strong intermolecular interaction, which hinders the lifting of Ag atoms out of the normal Ag(111) plane. This conclusion is further supported by our benchmark calculations involving fully deprotonated trimesic acid (TMA), which lacks attractive intermolecular interactions. In the case of TMA, silver atoms with three Ag-O bonds were lifted up, in line with previous works showing silver clusters in the molecular layer.[57] However, the diffraction model of the δ-BTB layer excludes such scenarios due to steric reasons: in the case of BTB, carboxyl groups are too far away to form three-fold Ag sites, and the molecular unit cell is too small to accommodate any silver adatom/cluster.

First, we will evaluate the preferred molecular phase in the case of the fully covered surface. There are two main contributions that decrease the free energy of the system: molecule-substrate bonding and intermolecular bonding. The computed energies per unit area reveal that the δ-BTB layer has by ~15 meV/Å$^2$ lower free energy per unit area than the mixed phase, i.e., the δ-BTB layer is more stable. This energy preference is elucidated by relatively strong Ag-O bonds, with a calculated binding energy of -1.7 eV, and supplemented by the contribution of attractive intermolecular interactions that stabilize the δ-BTB structure by an additional 0.8 eV per molecule. The strong attachment to the substrate results in the preference of BTB adsorption over the physisorbed pentacene. Hence, the complete δ-BTB layer shows a weak thermodynamic preference over the mixed phase.

Now, we will address the submonolayer coverages. The decisive parameter is still the surface free energy per unit area. However, in this case, there is a free substrate to accommodate all the adsorbed molecules irrespective of their bonding strength to the substrate. Since we are not restricted to the available surface area, the energy per molecule can be used to assess the preference for forming either pure or mixed phases. Our results show that the total adsorption energy per pentacene-BTB pair is 90 meV (PBE-D3) or 10 meV (optB86b) lower for the intermixed phase compared to the separate phases. However, the calculated stability is affected by imposed strain and the restriction to periodically repeating molecules that retain energetically unfavorable positions. To assess the validity of the results for the mixed structure, we have also computed its stability using modified supercells of similar dimensions but with different orientations with respect to the substrate, as shown in Supporting Information, Section 10. In this case, the highest stability achieved favors the mixed phase by 180 meV (PBE-D3) and 150 meV (optB86b) per one pentacene-BTB pair. These values present a lower limit for the stability of the mixed phase compared to the separate counterparts. In summary, these results point to the thermodynamic stability of the pentacene–BTB mixed phase for submonolayer coverages, which is consistent with experimental observations.

In the next step, we evaluate the kinetic barrier for breaking the compact δ-BTB layer. Due to the robust Ag–O bonds linking the BTB molecules to the silver substrate, the most likely scenario of disrupting the δ-BTB layer is to re-protonate the carboxyl groups, thus weakening their bonds to the surface, allowing their subsequent detachment from the surface. The deprotonated state is favored for a flat-laying BTB molecule, whereas the protonated carboxyl group is preferred for the BTB molecule detached from the surface. In detail, for a detached BTB, there is a 1.8 eV free energy preference for the protonated carboxyl group compared with the deprotonated group and 1/2 of $H_2$ molecule, taking into account the chemical potential of molecular hydrogen under conditions routinely reached during our experiments (-1.07 eV at 25 °C, $2\times10^{-10}$ mbar). On the contrary, for the flat laying BTB molecule, the formation of the O–H bond from molecular hydrogen is not favored; the free energy is by 0.2 eV higher compared with the molecular hydrogen under UHV conditions as the proximity of the silver substrate weakens the O–H bond. Therefore, the most probable way to disrupt the δ-BTB layer involves re-protonation of one of the carboxylic groups and its separation from the surface, resulting in a standing-up BTB configuration with the other two carboxylate groups attached to the substrate.

To estimate the energy barrier for opening the compact δ-BTB layer, one BTB molecule in the 2 × 2 supercell was arranged in the standing-up configuration, the lifted carboxylate group was protonated by additional hydrogen, and the whole structure was allowed to relax back to the flat-lying configuration. Figure 8 shows this process as a function of angle $α$ between the $z$-axis and a normal vector of the plane, which intersects the central phenyl ring. The detachment is composed of two modes: First, the non-linear up to 27° and the total energy difference between two limiting configurations of 0.87 eV; within this interval, the attractive intermolecular and molecule–substrate interactions are broken. The second mode shows a linear trend with an energy step of 18 meV per 1 degree. This behavior holds up to 70°, in which the total energy difference is estimated to be 1.7 eV. Initial and final structures are provided in the Supporting Information, Section 10. On the Ag(111) surface, the activation energy for the dissociation of hydrogen molecules amounts to 1.3 eV,[58] which is significantly larger than the barrier of 0.87 eV for the layer opening. This makes the hydrogen dissociation the rate-limiting step and the δ-BTB layer also kinetically stable at room temperature.

**Table 1.:** Calculated adsorption energies per molecule ($E_{ads}$) and energies per unit area ($\gamma$) for pentacene, deprotonated BTB (δ-BTB), and intermixed pentacene–BTB layer, using PBE-D3 and optB86 functionals. $E_{ads}$ for the intermixed phases is given for a pair comprising one BTB and one pentacene molecule, giving higher stability than pure molecular counterparts, i.e., a sum of the first two rows in a column.

| Molecular layer | $E_{ads}$ (eV) | | $\gamma$ (meV/Å$^2$) | |
|---|---|---|---|---|
| | PBE-D3 | optB86b | PBE-D3 | optB86b |
| Pentacene | -2.60 | -2.35 | -20.0 | -18.0 |
| δ-BTB | -9.36 | -9.44 | -61.7 | -62.0 |
| Intermixed (from exp. diffraction) | -12.05* | -11.79* | -42.8 | -41.7 |
| Intermixed (most stable) | -12.14* | -11.94* | -47.3 | -46.4 |

*per pentacene-BTB pair

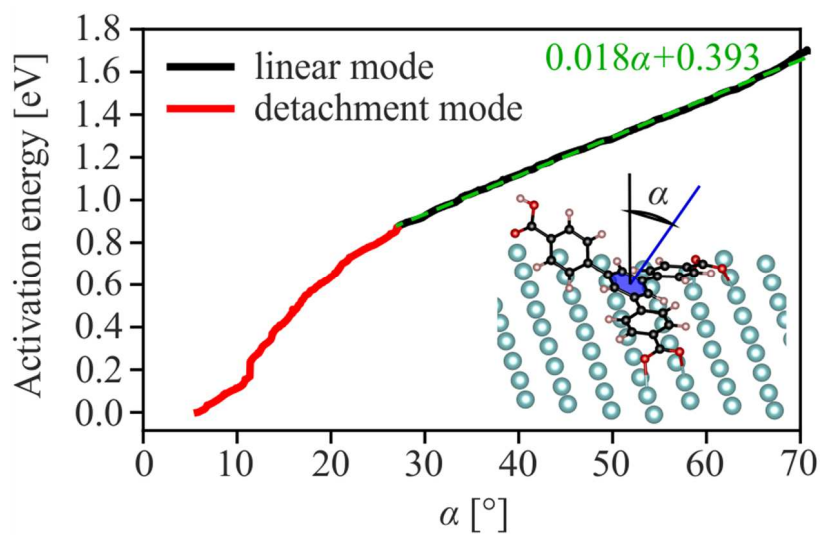

**Figure 8:** Detachment of one singly-protonated BTB molecule from the δ-BTB layer. For clarity, only the molecule being detached is shown. The detachment process is described as a function of the angle $\alpha$ between the *z*-axis and a normal vector of the plane that intersects the central phenyl ring (marked as blue in the inset). This process is composed of a non-linear mode up to 27° and 0.87 eV (red line). Above 27°, the trend is linear up to 70° with an energy step of 18 meV per 1 degree (black line).

**Discussion of the origin of the robustness of the compact δ-BTB layer**

Our experimental data and DFT calculations show the thermodynamic preference for the formation of mixed δ-BTB-pentacene phases. However, at the full coverage, the δ-BTB layer becomes preferred. This seemingly contradictory statement comes from the strong binding of carboxylate groups to the silver substrate, which defines the molecular structure. Hence, the other effects can take place only if all BTB molecules are bound to the substrate. Thus, for submonolayer coverages, there is a free area to satisfy the stability condition for the formation of the mixed pentacene–BTB phases, which are formed in the presence of supercritical[59] pentacene concentration.

The compact δ-BTB layer can be obtained by depositing >1 ML of BTB and subsequent sample annealing at the specific temperature. The excessive BTB desorbs from the surface, resulting in a compact δ-BTB layer without remaining BTB in the second layer. In contrast, obtaining the full layer of the BDA molecules (previous studies) was challenging as they display significant desorption from the first layer at temperatures close to full deprotonation. On Ag(111), the maximum coverage of the fully deprotonated BDA phase was around 50 %, and on Ag(100) it was between 90 – 95 %.

In the formation of the compact layer of deprotonated carboxylic acid molecules, the capability of filling the residual open sites is essential. This can be done by filling the gaps with molecules from the second layer. In the case of BTB, there are enough molecules in the second layer as their desorption temperature from the second BTB layer is close to the temperature required for their full deprotonation. On the contrary, the BDA molecules desorb from the second layer between 75 and 100 °C,[34] i.e., ~100 °C below the temperature required for the full deprotonation (180–200 °C)[49]. Hence, in the BDA case, there are no BDA molecules left to fill the gaps formed during the formation of the fully deprotonated phase.

The other essential feature is eliminating weak points in the layer that are generally localized at domain boundaries and step edges. Our STM images clearly show a favorable matching of molecular domains at their boundary so that they show minimum defects. The alignment of molecules at the step edges is generally challenging. However, the δ-BTB phase extends over several terraces, seamlessly growing over the step edges. The structure of BTB at the step edges is uniform along them, including the special shape of kink sites. The only possibility to achieve the perfect alignment of step edges with the growing molecular phase is to adjust their morphology to fit the growing molecular phase. The extended growth of 1D molecular chains,[60,61] 2D molecular phases,[62,63] or 2D metal–organic frameworks[64] featuring carboxylic[60–62,64] or cyano[63] groups was reported. This behavior was ascribed to the robust intermolecular bonding,[60,61,63] flexibility of the molecular backbone,[60,61] and adaptation of the substrate-step-edge shape and the terrace widths[62,64]. In our case, two contributions, i.e., molecular flexibility and changing the shape of the step edges, are important. Reshaping the substrate thus should be taken as an inherent part of the self-assembly of δ-BTB molecules, essential for the growth of the compact δ-BTB layer.

**Conclusions**

In this paper, we have introduced the robust dipolar layers based on the deprotonated tricarboxylic acid polyphenylene molecule, δ-BTB. The δ-BTB layer prepared on both low energy facets of silver, Ag(111) and Ag(100), cover the entire sample surface. This is enabled by the active sculpting of the surface by the BTB molecule, i.e., adjusting the position and shape of the step edges to fit the molecular layer precisely to the terraces and enable its seamless growth over the step edges. Similarly, the BTB molecules show a favorable matching at domain boundaries, which display a very low number of defects. The layer possesses the out-of-plane electric dipole that shifts the energy level of molecular layers deposited above, e.g., organic semiconductors, and, thus, can be employed as a monolayer thick charge injection layer for efficient charge injection from a metal electrode to an organic semiconductor film.

We have demonstrated the robustness of the layer towards organic semiconductors deposited on top of the compact δ-BTB layer and subsequent annealing at temperatures higher than typical processing temperatures of OS devices. The stability of the δ-BTB layer is given by the fact that it represents the energetically most favorable configuration. Even if one of the groups was protonated (e.g., under operational conditions), there still would be a considerable kinetic barrier associated with the disruption of the layer that would result in a possible loss of functionality.

Our work thus provides a robust molecular layer that can be employed as a wetting layer of subsequent growth of functional molecular films while it allows the control of substrate work function by adjusting the dipole density via the design of the molecular precursor. Our theoretical insights define the design criteria that ensure the robustness of the layer and allow the engineering of the layers with tailored functionality.

**Methods**

The experiments, including STM, LEEM, and XPS, were carried out in a complex ultrahigh vacuum (UHV) system at the CEITEC Nano Research Infrastructure, CZ. The system comprises several UHV chambers connected via a UHV transfer line. The investigated samples were prepared and analyzed in distinct UHV chambers, between which the samples were transferred through a UHV transfer line (base pressure $2\times10^{-10}$ mbar). During the transfers (60–150 s), the pressure increased to $\sim 2\times10^{-9}$ mbar but quickly recovered to the base level when the movement ceased.

**Sample Preparation.** The Ag(111) and Ag(100) single crystals (SPL) were cleaned by repeated cycles of $Ar^+$ sputtering and annealing at 520 °C, followed by slow cooling (< 1 K/s) to room temperature in the "Preparation Chamber" (base pressure of $2\times10^{-10}$ mbar). BTB molecules were subsequently deposited in an adjacent "Deposition Chamber" either by (i) a near-ambient effusion cell (Createc) from an oil-heated crucible held at 208°C or (ii) a Standard Effusion Cell (WEZ40, MBE Komponenten) from a resistively heated alumina crucible also held at 208°C on the sample held at room temperature. Pentacene, HM-TP, and HAT-CN molecules were all deposited from an Organic Material Evaporator Quad Cell (OEZ40, MBE Komponenten) from a resistively heated quartz crucible at temperatures 173 °C for pentacene, 170 °C for HM-TP, and 217 °C for HAT-CN on the samples held at room temperature. All mentioned molecular powders were purchased from Sigma−Aldrich (> 97% purity) and deposited after thorough degassing under UHV conditions at a pressure lower than $5\times10^{-9}$ mbar. The complete monolayer (ML) of BTB was obtained after 30 min deposition. We define 1 ML of BTB as a completely covered surface with the fully deprotonated phase, i.e., 0.045 BTB molecules per substrate unit cell area for Ag(111) and 0.048 for Ag(100). 1 ML of pentacene is defined as full coverage of pentacene forming p(6×3) superstructure on Ag(111) surface,[65] i.e., 0.056 pentacene molecule per substrate atom.

**Sample Analysis. Scanning Tunneling Microscopy (STM)** images were recorded with a commercial system Aarhus 150 (Specs) equipped with a KolibriSensor and a base pressure of $1\times10^{-10}$ mbar. Images from the STM measurements were acquired at room temperature in constant current mode; the sample bias voltage was set between 1.0 V and 1.4 V, and the tunneling current was set to 50 pA, if not otherwise indicated in the images. The STM images (and diffraction patterns, see LEEM below) were rotated to align the horizontal direction of the images with the main crystallographic direction of the Ag(111) substrate. The drift distortion in the images was corrected by transforming the image to match the unit cell determined in STM with the one obtained from diffraction in LEEM. **Low-Energy Electron Microscopy/Diffraction (LEEM/LEED)** experiments were performed in a Specs FE-LEEM P90 instrument with a base pressure of $2\times10^{-10}$ mbar. Bright-field images were formed by electrons collected from the (0,0) diffracted beam. Diffraction patterns were obtained from a $15\times10$ μm$^2$ surface area, and microdiffraction was done with a 185 nm e-beam spot size on the sample. The diffraction patterns were modeled using ProLEED Studio;[66] a local congruence approach[46,47] was employed to match the modeled diffraction patterns with the measured ones. **X-ray Photoelectron Spectroscopy (XPS)** was performed on a Specs system utilizing a Phoibos 150 spectrometer. All measurements employed a non-monochromatized X-ray source with Mg Kα radiation ($h\nu$=1253.6 eV), emission angle 0°, and the high magnification mode with the iris aperture set to 15 mm. Detailed spectra were acquired in medium magnification mode, 20 eV pass energy, 0.1 eV energy step for C 1s and 0.05 eV for O 1s region; for each spectrum 40 sweeps with 0.1 s dwell time were summed. The total resolution (accounting for analyzer and excitation radiation contributions) was 800 meV. The photoelectron spectra were subsequently fitted by Voigt profiles after a Shirley or (Shirley+linear) background subtraction. **Synchrotron radiation photoelectron spectroscopy** was performed at the Materials Science Beamline at the Elettra synchrotron light source in Trieste, IT. We used excitation energies of 420, 510, and 670 eV for the C 1s, Ag 3d$_{5/2}$, and O 1s peaks, respectively. Detailed spectra were acquired in medium area lens mode using 10 eV pass energy integrating 2 (Ag 3d), 5 (C 1s), or 20 (O 1s) sweeps; the dwell time was set to 0.1 s, and the energy step size to 0.05 eV. The total resolution was in the range of 300–550 meV. Peak positions were corrected with respect to the measured Fermi edge of the Ag substrate and intensity for the photo-current of a gold mesh placed in the beamline. The temperature was read by a K-type thermocouple attached to the bottom side of the sample plate (Ta, thickness 0.1 mm). The WF was determined from the position of the secondary electron cut-off as described earlier.[34]

**Computational details:** All density functional theory calculations are carried out using the Vienna *ab initio* simulation package (VASP).[67] For silver, carbon, hydrogen, and oxygen, 11 valence electrons, 4 valence electrons, 1 valence electron, and 6 valence electrons, respectively, are described with a plane-wave basis set with an energy cut-off set to 450 eV while all core electrons are treated by the projector augmented wave method (PAW).[68] The exchange-correlation energy is described by the PBE-D3 functional[69,70] and benchmarked with the non-local optB86b van-der-Waals functional,[71] which shows no qualitative differences in the presented results. Structural and electronic calculations are performed in two steps: first, for all structural relaxations, we use a sparser *k*-points grid: 3×3×1 Gamma-centered Monkhorst-pack grid[72] for the δ-BTB phase and gamma-only calculations for other structures; and the geometry optimization is stopped when all residual forces acting on ions are smaller than 0.01 eV/Å. Second, electronic structure calculations are performed as single-point calculations starting from optimized structures with a denser sampling of the Brillouin zone with a grid density larger than 60 *k*-points per Å$^{-1}$. The Ag(111) surface is modeled with a 5-layered slab, and all calculations account for dipole corrections to the potential and energy.

## ASSOCIATED CONTENT

Supplemental Material: (1) Synchrotron radiation photoelectron spectroscopy analysis of the BTB deprotonation on Ag(111) and Ag(100); (2) STM images of the as-deposited phase and decarboxylated polymeric network; (3) Compact δ-BTB layer and pentacene–BTB mixed phases on Ag(100); (4) Laboratory XPS analysis of the compact δ-BTB layer; (5) Supplementary STM figures of δ-BTB on Ag(111); (6) Comparison of δ-BTB unit cell on Ag(100) and Ag(111); (7) Stable overlayers of HM-TP and HAT-CN; (8) The wheel-like pentacene-BTB mixed phase on Ag(111); (9) Gas-phase reference for calculated adsorption energies; and (10) Supplementary DFT structures of molecular monolayers on Ag(111) substrate.

## AUTHOR INFORMATION

**Corresponding Author**

* E-mail: cechal@fme.vutbr.cz (J. Č.)

**Author Contributions**

T.K. prepared the samples, performed laboratory (LEEM, STM, and XPS) measurements on Ag(111), and wrote the initial manuscript. V.S. prepared the samples and performed laboratory measurements on Ag(100). V.S., P.P., T.S., M.B., and J.Č. performed the synchrotron radiation experiments. P.P., together with T.K. and V.S., evaluated LEEM/LEED data. M.B., together with T.K., evaluated STM data. J.P. performed the DFT calculations. J.Č. evaluated the XPS and synchrotron data. P.P. and J.Č. coordinated the research. J.Č. wrote the manuscript. All authors discussed the results and contributed to the final manuscript.

## COMPETING INTERESTS

The authors declare no competing interests.

## ACKNOWLEDGMENT

This research has been supported by GAČR, project No. 23-08001S. We acknowledge CzechNanoLab Research Infrastructure (LM2023051) and e-INFRA CZ (ID:90254) supported by MEYS CR, and the CERIC-ERIC Consortium for access to experimental/computational facilities and financial support. J.P. and M.B. were supported by the ESF under project CZ.02.01.01/00/22_010/0002552.

## DATA AVAILABILITY

The data that support the findings of this study are available from the corresponding author upon reasonable request.

REFERENCES

(1) Huang, Y.; Hsiang, E.-L.; Deng, M.-Y.; Wu, S.-T. Mini-LED, Micro-LED and OLED Displays: Present Status and Future Perspectives. *Light Sci. Appl.* **2020**, *9* (1), 105. https://doi.org/10.1038/s41377-020-0341-9.

(2) Wang, S.; Zhang, H.; Zhang, B.; Xie, Z.; Wong, W. Towards High-Power-Efficiency Solution-Processed OLEDs: Material and Device Perspectives. *Mater. Sci. Eng. R Reports* **2020**, *140* (January), 100547. https://doi.org/10.1016/j.mser.2020.100547.

(3) Zou, S.; Shen, Y.; Xie, F.; Chen, J.; Li, Y.; Tang, J.-X. Recent Advances in Organic Light-Emitting Diodes: Toward Smart Lighting and Displays. *Mater. Chem. Front.* **2020**, *4* (3), 788–820. https://doi.org/10.1039/C9QM00716D.

(4) Park, S. K.; Kim, J. H.; Park, S. Y. Organic 2D Optoelectronic Crystals: Charge Transport, Emerging Functions, and Their Design Perspective. *Adv. Mater.* **2018**, *30*, 1704759. https://doi.org/10.1002/adma.201704759.

(5) Gao, J.; Wang, J.; Xu, C.; Hu, Z.; Ma, X.; Zhang, X.; Niu, L.; Zhang, J.; Zhang, F. A Critical Review on Efficient Thick-Film Organic Solar Cells. *Sol. RRL* **2020**, *4*, 2000364. https://doi.org/10.1002/solr.202000364.

(6) Yu, Y.; Ma, Q.; Ling, H.; Li, W.; Ju, R.; Bian, L.; Shi, N.; Qian, Y.; Yi, M.; Xie, L.; et al. Small-Molecule-Based Organic Field-Effect Transistor for Nonvolatile Memory and Artificial Synapse. *Adv. Funct. Mater.* **2019**, *29* (50), 1904602. https://doi.org/10.1002/adfm.201904602.

(7) Waldrip, M.; Jurchescu, O. D.; Gundlach, D. J.; Bittle, E. G. Contact Resistance in Organic Field-Effect Transistors: Conquering the Barrier. *Adv. Funct. Mater.* **2020**, *30*, 1904576. https://doi.org/10.1002/adfm.201904576.

(8) Klauk, H. Will We See Gigahertz Organic Transistors? *Adv. Electron. Mater.* **2018**, *4*, 1700474. https://doi.org/10.1002/aelm.201700474.

(9) Koch, N. Opportunities for Energy Level Tuning at Inorganic/Organic Semiconductor Interfaces. *Appl. Phys. Lett.* **2021**, *119*, 260501. https://doi.org/10.1063/5.0074963.

(10) Franco-Cañellas, A.; Duhm, S.; Gerlach, A.; Schreiber, F. Binding and Electronic Level Alignment of π -Conjugated Systems on Metals. *Reports Prog. Phys.* **2020**, *83* (6), 066501. https://doi.org/10.1088/1361-6633/ab7a42.

(11) Otero, R.; Vázquez de Parga, A. L.; Gallego, J. M. Electronic, Structural and Chemical Effects of Charge-Transfer at Organic/Inorganic Interfaces. *Surf. Sci. Rep.* **2017**, *72*, 105–145. https://doi.org/10.1016/j.surfrep.2017.03.001.

(12) Fahlman, M.; Fabiano, S.; Gueskine, V.; Simon, D.; Berggren, M.; Crispin, X. Interfaces in Organic Electronics. *Nat. Rev. Mater.* **2019**, *4* (10), 627–650. https://doi.org/10.1038/s41578-019-0127-y.

(13) Zojer, E.; Taucher, T. C.; Hofmann, O. T. The Impact of Dipolar Layers on the Electronic Properties of Organic/Inorganic Hybrid Interfaces. *Adv. Mater. Interfaces* **2019**, *6*, 1900581. https://doi.org/10.1002/admi.201900581.

(14) Goiri, E.; Borghetti, P.; El-Sayed, A.; Ortega, J. E.; de Oteyza, D. G. Multi-Component Organic Layers on Metal Substrates. *Adv. Mater.* **2016**, *28*, 1340–1368. https://doi.org/10.1002/adma.201503570.

(15) Chen, H.; Zhang, W.; Li, M.; He, G.; Guo, X. Interface Engineering in Organic Field-Effect Transistors: Principles, Applications, and Perspectives. *Chem. Rev.* **2020**, *120* (5), 2879–2949. https://doi.org/10.1021/acs.chemrev.9b00532.

(16) Lim, K.-G.; Ahn, S.; Lee, T.-W. Energy Level Alignment of Dipolar Interface Layer in Organic and Hybrid Perovskite Solar Cells. *J. Mater. Chem. C* **2018**, *6*, 2915–2924. https://doi.org/10.1039/C8TC00166A.

(17) Borchert, J. W.; Weitz, R. T.; Ludwigs, S.; Klauk, H. A Critical Outlook for the Pursuit of

Lower Contact Resistance in Organic Transistors. *Adv. Mater.* **2022**, *34* (2), 2104075. https://doi.org/10.1002/adma.202104075.

(18) Zojer, E.; Terfort, A.; Zharnikov, M. Concept of Embedded Dipoles as a Versatile Tool for Surface Engineering. *Acc. Chem. Res.* **2022**, *55*, 1857–1867. https://doi.org/10.1021/acs.accounts.2c00173.

(19) Casalini, S.; Bortolotti, C. A.; Leonardi, F.; Biscarini, F. Self-Assembled Monolayers in Organic Electronics. *Chem. Soc. Rev.* **2017**, *46*, 40–71. https://doi.org/10.1039/C6CS00509H.

(20) Stoliar, P.; Kshirsagar, R.; Massi, M.; Annibale, P.; Albonetti, C.; de Leeuw, D. M.; Biscarini, F. Charge Injection Across Self-Assembly Monolayers in Organic Field-Effect Transistors: Odd−Even Effects. *J. Am. Chem. Soc.* **2007**, *129*, 6477–6484. https://doi.org/10.1021/ja069235m.

(21) Kovalchuk, A.; Abu-Husein, T.; Fracasso, D.; Egger, D. A.; Zojer, E.; Zharnikov, M.; Terfort, A.; Chiechi, R. C. Transition Voltages Respond to Synthetic Reorientation of Embedded Dipoles in Self-Assembled Monolayers. *Chem. Sci.* **2016**, *7*, 781–787. https://doi.org/10.1039/C5SC03097H.

(22) Zhang, J. L.; Ye, X.; Gu, C.; Han, C.; Sun, S.; Wang, L.; Chen, W. Non-Covalent Interaction Controlled 2D Organic Semiconductor Films: Molecular Self-Assembly, Electronic and Optical Properties, and Electronic Devices. *Surf. Sci. Rep.* **2020**, *75*, 100481. https://doi.org/10.1016/j.surfrep.2020.100481.

(23) Amsalem, P.; Wilke, A.; Frisch, J.; Niederhausen, J.; Vollmer, A.; Rieger, R.; Müllen, K.; Rabe, J. P.; Koch, N. Interlayer Molecular Diffusion and Thermodynamic Equilibrium in Organic Heterostructures on a Metal Electrode. *J. Appl. Phys.* **2011**, *110* (11), 113709. https://doi.org/10.1063/1.3662878.

(24) Sun, L.; Liu, C.; Queteschiner, D.; Weidlinger, G.; Zeppenfeld, P. Layer Inversion in Organic

Heterostructures. *Phys. Chem. Chem. Phys.* **2011**, *13* (29), 13382. https://doi.org/10.1039/c1cp21151j.

(25) Häming, M.; Greif, M.; Sauer, C.; Schöll, A.; Reinert, F. Electronic Structure of Ultrathin Heteromolecular Organic-Metal Interfaces: SnPc/PTCDA/Ag(111) and SnPc/Ag(111). *Phys. Rev. B* **2010**, *82*, 235432. https://doi.org/10.1103/PhysRevB.82.235432.

(26) Egger, D. A.; Ruiz, V. G.; Saidi, W. A.; Bučko, T.; Tkatchenko, A.; Zojer, E. Understanding Structure and Bonding of Multilayered Metal–Organic Nanostructures. *J. Phys. Chem. C* **2013**, *117*, 3055–3061. https://doi.org/10.1021/jp309943k.

(27) Gallego, J. M.; Ecija, D.; Martín, N.; Otero, R.; Miranda, R. An STM Study of Molecular Exchange Processes in Organic Thin Film Growth. *Chem. Commun.* **2014**, *50* (69), 9954–9957. https://doi.org/10.1039/C4CC03656E.

(28) Stadtmüller, B.; Schröder, S.; Kumpf, C. Heteromolecular Metal-Organic Interfaces: Electronic and Structural Fingerprints of Chemical Bonding. *J. Electron Spectros. Relat. Phenomena* **2015**, *204*, 80–91. https://doi.org/10.1016/j.elspec.2015.03.003.

(29) Borghetti, P.; de Oteyza, D. G.; Rogero, C.; Goiri, E.; Verdini, A.; Cossaro, A.; Floreano, L.; Ortega, J. E. Molecular-Level Realignment in Donor–Acceptor Bilayer Blends on Metals. *J. Phys. Chem. C* **2016**, *120*, 5997–6005. https://doi.org/10.1021/acs.jpcc.5b11373.

(30) Thussing, S.; Jakob, P. Thermal Stability and Interlayer Exchange Processes in Heterolayers of CuPc and PTCDA on Ag(111). *J. Phys. Chem. C* **2017**, *121*, 13680–13691. https://doi.org/10.1021/acs.jpcc.7b02377.

(31) Wang, Q.; Franco-Cañellas, A.; Ji, P.; Bürker, C.; Wang, R.-B.; Broch, K.; Thakur, P. K.; Lee, T.-L.; Zhang, H.; Gerlach, A.; et al. Bilayer Formation vs Molecular Exchange in Organic Heterostructures: Strong Impact of Subtle Changes in Molecular Structure. *J. Phys. Chem. C* **2018**, *122*, 9480–9490. https://doi.org/10.1021/acs.jpcc.8b01529.

(32) Lerch, A.; Zimmermann, J. E.; Namgalies, A.; Stallberg, K.; Höfer, U. Two-Photon Photoemission Spectroscopy of Unoccupied Electronic States at CuPc/PTCDA/Ag(1 1 1) Interfaces. *J. Phys. Condens. Matter* **2018**, *30*, 494001. https://doi.org/10.1088/1361-648X/aaec53.

(33) Wang, Q.; Franco-Cañellas, A.; Yang, J.; Hausch, J.; Struzek, S.; Chen, M.; Thakur, P. K.; Gerlach, A.; Duhm, S.; Schreiber, F. Heteromolecular Bilayers on a Weakly Interacting Substrate: Physisorptive Bonding and Molecular Distortions of Copper–Hexadecafluorophthalocyanine. *ACS Appl. Mater. Interfaces* **2020**, *12*, 14542–14551. https://doi.org/10.1021/acsami.9b22812.

(34) Stará, V.; Procházka, P.; Planer, J.; Shahsavar, A.; Makoveev, A. O.; Skála, T.; Blatnik, M.; Čechal, J. Tunable Energy-Level Alignment in Multilayers of Carboxylic Acids on Silver. *Phys. Rev. Appl.* **2022**, *18* (4), 044048. https://doi.org/10.1103/PhysRevApplied.18.044048.

(35) Schlotter, N. E.; Porter, M. D.; Bright, T. B.; Allara, D. L. Formation and Structure of a Spontaneously Adsorbed Monolayer of Arachidic on Silver. *Chem. Phys. Lett.* **1986**, *132*, 93–98. https://doi.org/10.1016/0009-2614(86)80702-3.

(36) Tao, Y. T. Structural Comparison of Self-Assembled Monolayers of n-Alkanoic Acids on the Surfaces of Silver, Copper, and Aluminum. *J. Am. Chem. Soc.* **1993**, *115*, 4350–4358. https://doi.org/10.1021/ja00063a062.

(37) Tao, Y. T.; Huang, C. Y.; Chiou, D. R.; Chen, L. J. Infrared and Atomic Force Microscopy Imaging Study of the Reorganization of Self-Assembled Monolayers of Carboxylic Acids on Silver Surface. *Langmuir* **2002**, *18*, 8400–8406. https://doi.org/10.1021/la025805u.

(38) Aitchison, H.; Lu, H.; Hogan, S. W. L.; Früchtl, H.; Cebula, I.; Zharnikov, M.; Buck, M. Self-Assembled Monolayers of Oligophenylenecarboxylic Acids on Silver Formed at the Liquid–Solid Interface. *Langmuir* **2016**, *32*, 9397–9409. https://doi.org/10.1021/acs.langmuir.6b01773.

(39) Krzykawska, A.; Ossowski, J.; Żaba, T.; Cyganik, P. Binding Groups for Highly Ordered SAM Formation: Carboxylic versus Thiol. *Chem. Commun.* **2017**, *53*, 5748–5751. https://doi.org/10.1039/C7CC01939D.

(40) Krzykawska, A.; Szwed, M.; Ossowski, J.; Cyganik, P. Odd–Even Effect in Molecular Packing of Self-Assembled Monolayers of Biphenyl-Substituted Fatty Acid on Ag(111). *J. Phys. Chem. C* **2018**, *122*, 919–928. https://doi.org/10.1021/acs.jpcc.7b10806.

(41) Fahlman, M.; Fabiano, S.; Gueskine, V.; Simon, D.; Berggren, M.; Crispin, X. Interfaces in Organic Electronics. *Nat. Rev. Mater.* **2019**, *4* (10), 627–650. https://doi.org/10.1038/s41578-019-0127-y.

(42) Goronzy, D. P.; Ebrahimi, M.; Rosei, F.; Arramel; Fang, Y.; De Feyter, S.; Tait, S. L.; Wang, C.; Beton, P. H.; Wee, A. T. S.; et al. Supramolecular Assemblies on Surfaces: Nanopatterning, Functionality, and Reactivity. *ACS Nano* **2018**, *12*, 7445–7481. https://doi.org/10.1021/acsnano.8b03513.

(43) Deimel, P. S.; Feulner, P.; Barth, J. V.; Allegretti, F. Spatial Decoupling of Macrocyclic Metal–Organic Complexes from a Metal Support: A 4-Fluorothiophenol Self-Assembled Monolayer as a Thermally Removable Spacer. *Phys. Chem. Chem. Phys.* **2019**, *21*, 10992–11003. https://doi.org/10.1039/C9CP01583C.

(44) Widdascheck, F.; Bischof, D.; Witte, G. Engineering of Printable and Air-Stable Silver Electrodes with High Work Function Using Contact Primer Layer: From Organometallic Interphases to Sharp Interfaces. *Adv. Funct. Mater.* **2021**, *31* (49), 2106687. https://doi.org/10.1002/adfm.202106687.

(45) Stadtmüller, B.; Sueyoshi, T.; Kichin, G.; Kröger, I.; Soubatch, S.; Temirov, R.; Tautz, F. S.; Kumpf, C. Commensurate Registry and Chemisorption at a Hetero-Organic Interface. *Phys. Rev. Lett.* **2012**, *108* (10), 1–5. https://doi.org/10.1103/PhysRevLett.108.106103.

(46) Procházka, P.; Gosalvez, M. A.; Kormoš, L.; De La Torre, B.; Gallardo, A.; Alberdi-Rodriguez, J.; Chutora, T.; Makoveev, A. O.; Shahsavar, A.; Arnau, A.; et al. Multiscale Analysis of Phase Transformations in Self-Assembled Layers of 4,4′-Biphenyl Dicarboxylic Acid on the Ag(001) Surface. *ACS Nano* **2020**, *14*, 7269–7279. https://doi.org/10.1021/acsnano.0c02491.

(47) Kormoš, L.; Procházka, P.; Makoveev, A. O.; Čechal, J. Complex K-Uniform Tilings by a Simple Bitopic Precursor Self-Assembled on Ag(001) Surface. *Nat. Commun.* **2020**, *11*, 1856. https://doi.org/10.1038/s41467-020-15727-6.

(48) Procházka, P.; Kormoš, L.; Shahsavar, A.; Stará, V.; Makoveev, A. O.; Skála, T.; Blatnik, M.; Čechal, J. Phase Transformations in a Complete Monolayer of 4,4′-Biphenyl-Dicarboxylic Acid on Ag(0 0 1). *Appl. Surf. Sci.* **2021**, *547*, 149115. https://doi.org/10.1016/j.apsusc.2021.149115.

(49) Makoveev, A. O.; Procházka, P.; Blatnik, M.; Kormoš, L.; Skála, T.; Čechal, J. Role of Phase Stabilization and Surface Orientation in 4,4′-Biphenyl-Dicarboxylic Acid Self-Assembly and Transformation on Silver Substrates. *J. Phys. Chem. C* **2022**, *126*, 9989–9997. https://doi.org/10.1021/acs.jpcc.2c02538.

(50) Makoveev, A.; Procházka, P.; Shahsavar, A.; Kormoš, L.; Krajňák, T.; Stará, V.; Čechal, J. Kinetic Control of Self-Assembly Using a Low-Energy Electron Beam. *Appl. Surf. Sci.* **2022**, *600*, 154106. https://doi.org/10.1016/j.apsusc.2022.154106.

(51) Fratini, S.; Nikolka, M.; Salleo, A.; Schweicher, G.; Sirringhaus, H. Charge Transport in High-Mobility Conjugated Polymers and Molecular Semiconductors. *Nat. Mater.* **2020**, *19*, 491–502. https://doi.org/10.1038/s41563-020-0647-2.

(52) MacLeod, J. Design and Construction of On-Surface Molecular Nanoarchitectures: Lessons and Trends from Trimesic Acid and Other Small Carboxlyated Building Blocks. *J. Phys. D* **2020**, *53*, 043002. https://doi.org/10.1088/1361-6463/ab4c4d.

(53) Kormoš, L.; Procházka, P.; Šikola, T.; Čechal, J. Molecular Passivation of Substrate Step Edges as Origin of Unusual Growth Behavior of 4,4′-Biphenyl Dicarboxylic Acid on Cu(001). *J. Phys. Chem. C* **2018**, *122*, 2815–2820.

(54) Derry, G. N.; Kern, M. E.; Worth, E. H. Recommended Values of Clean Metal Surface Work Functions. *J. Vac. Sci. Technol. A Vacuum, Surfaces, Film.* **2015**, *33*, 060801. https://doi.org/10.1116/1.4934685.

(55) El-Sayed, A.; Borghetti, P.; Goiri, E.; Rogero, C.; Floreano, L.; Lovat, G.; Mowbray, D. J.; Cabellos, J. L.; Wakayama, Y.; Rubio, A.; et al. Understanding Energy-Level Alignment in Donor–Acceptor/Metal Interfaces from Core-Level Shifts. *ACS Nano* **2013**, *7*, 6914–6920. https://doi.org/10.1021/nn4020888.

(56) Ruben, M.; Payer, D.; Landa, A.; Comisso, A.; Gattinoni, C.; Lin, N.; Collin, J.-P.; Sauvage, J.-P.; De Vita, A.; Kern, K. 2D Supramolecular Assemblies of Benzene-1,3,5-Triyl-Tribenzoic Acid: Temperature-Induced Phase Transformations and Hierarchical Organization with Macrocyclic Molecules. *J. Am. Chem. Soc.* **2006**, *128* (49), 15644–15651. https://doi.org/10.1021/ja063601k.

(57) Svane, K. L.; Baviloliaei, M. S.; Hammer, B.; Diekhöner, L. An Extended Chiral Surface Coordination Network Based on Ag7-Clusters. *J. Chem. Phys.* **2018**, *149*. https://doi.org/10.1063/1.5051510.

(58) Mohammad, A. B.; Hwa Lim, K.; Yudanov, I. V.; Neyman, K. M.; Rösch, N. A Computational Study of H 2 Dissociation on Silver Surfaces: The Effect of Oxygen in the Added Row Structure of Ag(110). *Phys. Chem. Chem. Phys.* **2007**, *9*, 1247–1254. https://doi.org/10.1039/B616675J.

(59) Henneke, C.; Felter, J.; Schwarz, D.; Tautz, F. S.; Kumpf, C. Controlling the Growth of Multiple Ordered Heteromolecular Phases by Utilizing Intermolecular Repulsion. *Nat. Mater.* **2017**, *16*, 628–633. https://doi.org/10.1038/NMAT4858.

(60) Schnadt, J.; Rauls, E.; Xu, W.; Vang, R. T.; Knudsen, J.; Lagsgaard, E.; Li, Z.; Hammer, B.; Besenbacher, F. Extended One-Dimensional Supramolecular Assembly on a Stepped Surface. *Phys. Rev. Lett.* **2008**, *100* (4), 046103. https://doi.org/10.1103/PhysRevLett.100.046103.

(61) Schnadt, J.; Xu, W.; Vang, R. T.; Knudsen, J.; Li, Z.; Lægsgaard, E.; Besenbacher, F. Interplay of Adsorbate-Adsorbate and Adsorbate-Substrate Interactions in Self-Assembled Molecular Surface Nanostructures. *Nano Res.* **2010**, *3*, 459–471. https://doi.org/10.1007/s12274-010-0005-9.

(62) Pascual, J. I.; Barth, J. V.; Ceballos, G.; Trimarchi, G.; De Vita, A.; Kern, K.; Rust, H. P. Mesoscopic Chiral Reshaping of the Ag(110) Surface Induced by the Organic Molecule PVBA. *J. Chem. Phys.* **2004**, *120*, 11367–11370. https://doi.org/10.1063/1.1763836.

(63) Kim, J.-H.; Ribierre, J.-C.; Yang, Y. S.; Adachi, C.; Kawai, M.; Jung, J.; Fukushima, T.; Kim, Y. Seamless Growth of a Supramolecular Carpet. *Nat. Commun.* **2016**, *7*, 10653. https://doi.org/10.1038/ncomms10653.

(64) Kley, C. S.; Čechal, J.; Kumagai, T.; Schramm, F.; Ruben, M.; Stepanow, S.; Kern, K. Highly Adaptable Two-Dimensional Metal-Organic Coordination Networks on Metal Surfaces. *J. Am. Chem. Soc.* **2012**, *134*, 6072–6075. https://doi.org/10.1021/ja211749b.

(65) Mete, E.; Demiroğlu, İ.; Fatih Danışman, M.; Ellialtıoğlu, Ş. Pentacene Multilayers on Ag(111) Surface. *J. Phys. Chem. C* **2010**, *114*, 2724–2729. https://doi.org/10.1021/jp910703n.

(66) Procházka, P.; Čechal, J. ProLEED Studio : Software for Modeling Low-Energy Electron Diffraction Patterns. *J. Appl. Crystallogr.* **2024**, *57*, 187. https://doi.org/10.1107/S1600576723010312.

(67) Kresse, G.; Hafner, J. Ab Initio Molecular Dynamics for Liquid Metals. *Phys. Rev. B* **1993**, *47*, 558–561.

(68) Kresse, G.; Joubert, D. From Ultrasoft Pseudopotentials to the Projector Augmented-Wave

Method. *Phys. Rev. B* **1999**, *59*, 1758–1775. https://doi.org/10.1103/PhysRevB.59.1758.

(69) Perdew, J. P.; Burke, K.; Ernzerhof, M. Generalized Gradient Approximation Made Simple. *Phys. Rev. Lett.* **1996**, *77*, 3865–3868. https://doi.org/10.1103/PhysRevLett.77.3865.

(70) Grimme, S.; Antony, J.; Ehrlich, S.; Krieg, H. A Consistent and Accurate Ab Initio Parametrization of Density Functional Dispersion Correction (DFT-D) for the 94 Elements H-Pu. *J. Chem. Phys.* **2010**, *132*, 154104. https://doi.org/10.1063/1.3382344.

(71) Klimeš, J.; Bowler, D. R.; Michaelides, A. Van Der Waals Density Functionals Applied to Solids. *Phys. Rev. B* **2011**, *83*, 195131. https://doi.org/10.1103/PhysRevB.83.195131.

(72) Monkhorst, H. J.; Pack, J. D. Special Points for Brillouin-Zone Integrations. *Phys. Rev. B* **1976**, *13*, 5188–5192. https://doi.org/10.1103/PhysRevB.13.5188.

TOC Graphics

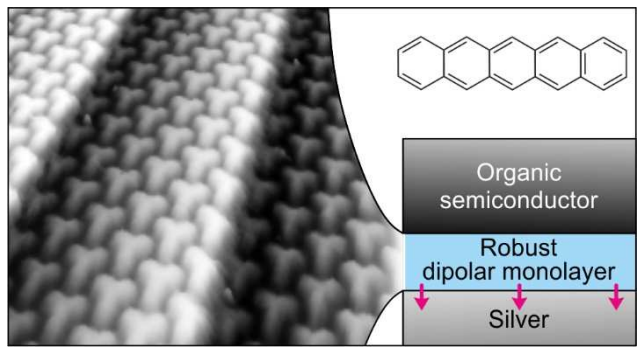

SUPPORTING INFORMATION

# Robust Dipolar Layers between Organic Semiconductors and Silver for Energy-Level Alignment

*Tomáš Krajňák,[1] Veronika Stará,[1] Pavel Procházka,[1] Jakub Planer,[1] Tomáš Skála,[2] Matthias Blatnik,[1] Jan Čechal[1,3]\**

[1] CEITEC - Central European Institute of Technology, Brno University of Technology, Purkyňova 123, 612 00 Brno, Czech Republic.

[2] Department of Surface and Plasma Science, Faculty of Mathematics and Physics, Charles University, V Holešovičkách 2, 180 00 Prague 8, Czech Republic.

[3] Institute of Physical Engineering, Brno University of Technology, Technická 2896/2, 616 69 Brno, Czech Republic.

\* E-mail: cechal@fme.vutbr.cz (J. Č.)

CONTENTS:

# 1. Synchrotron radiation photoelectron spectroscopy analysis of the BTB deprotonation on Ag(111) and Ag(100)

Contrary to our previous papers, where BDA molecules were employed, we could not unambiguously relate the measured spectra to well-defined molecular phases measured by LEEM. Hence, the XPS analysis is targeted primarily to obtain the degree of deprotonation at given annealing temperatures and relate it to simultaneously measured work function.

The C 1s spectra were measured after annealing up to a temperature of 280 °C. An example of spectra measured for as-deposited, intermediate, and fully deprotonated BTB phases is given in Figure S1 for both Ag(111) and Ag(100) substrates. The main component of C 1s spectra at ~285 eV can be associated with carbon atoms in phenyl rings and the small component at higher binding energies (287 – 289 eV) with carboxylic carbon. The fitting of the C 1s spectra is not straightforward due to the presence of shake-up satellites; however, detailed fits are unnecessary for this work. Hence, the C 1s spectra are presented without fitting in Figure S1.

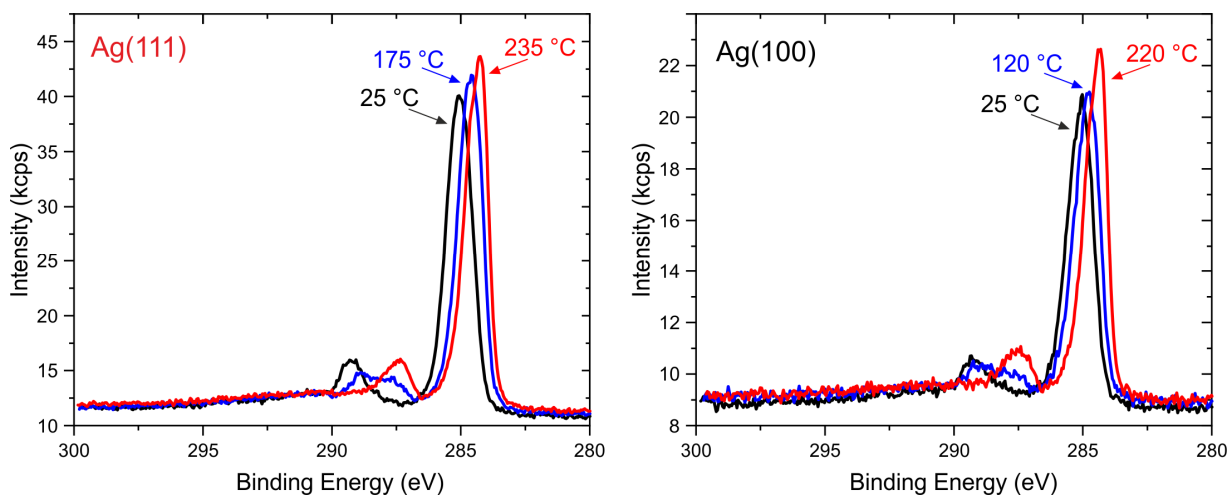

**Figure S1:** C 1s spectra of BTB layer measured by synchrotron radiation photoelectron spectroscopy for as-deposited sample (black), after annealing at the temperatures of 175 or 120 °C (blue), and 235 or 220 °C (red) on Ag(111) and Ag(100) surfaces.

**Fitting of the O 1s spectra**

The O 1s spectra were fitted by up to 5 Voigt components and a Shirley background, as shown in Figure S2; the peak fitting parameters are summarized in Tables S1 for Ag(111) and S2 for Ag(100). The O 1s spectrum of the as-deposited BTB can be fitted by two pairs of peaks: the first pair (O1) at 533.8 + 532.5 eV (light blue and blue in Figure S2) and the second pair (O2) at 532.0 + 531.1 eV (light green and green in Figure S2). The intensity ratio of these pairs was 2:1. Within the pair, the intensities of the components have a ratio of 1:1; the higher binding energy component can be associated with hydroxyl, and the lower binding energy component with carbonyl oxygen. For annealed samples, we have added a peak component (O3) associated with a deprotonated carboxyl group (red in Figure S2); a single component is due to the symmetric chemical state of both oxygen atoms. The degree of deprotonation of carboxyl groups was determined as a ratio of carboxylate (O3) to the total intensity of O 1s peak. For BDA (previous works), we introduced two additional components related to the intact carboxyl group bound to the deprotonated one; their position would be similar to the green components (O2). Therefore, in the analysis of BTB, we did not include them, which results in a generally imprecise fitting, which, however, gives the correct degree of deprotonation.

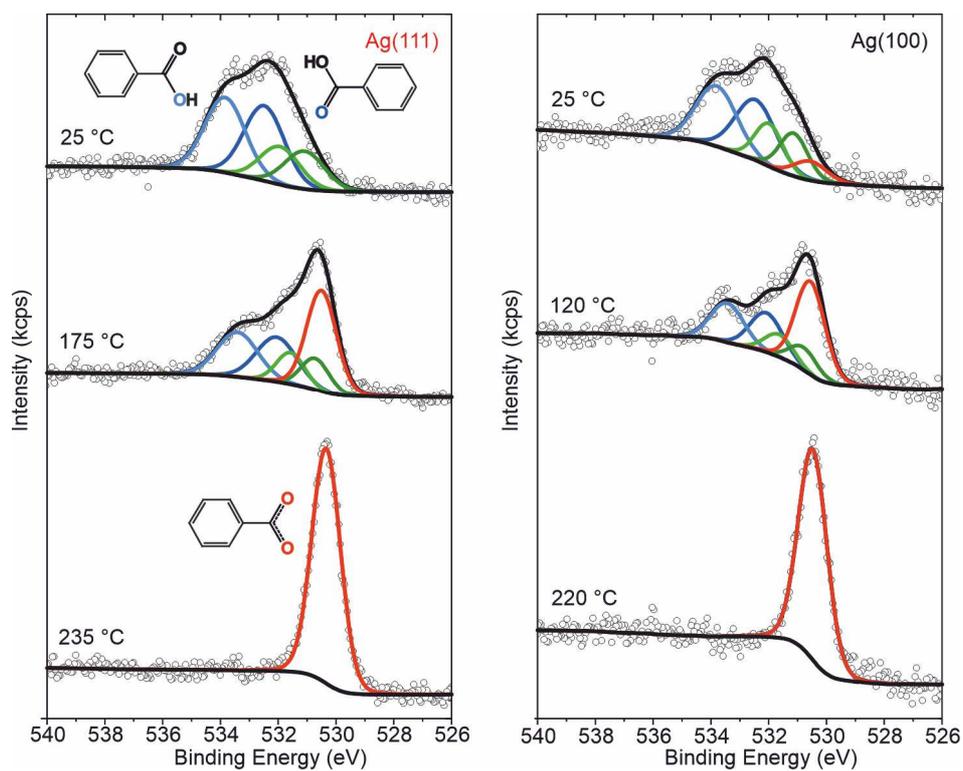

**Figure S2**: Example of fitting of the O 1s spectra measured by synchrotron radiation photoelectron spectroscopy for as-deposited sample (top row), partially deprotonated BTB annealed at the temperature of 175 or 120 °C (middle), and fully deprotonated BTB annealed at 235 or 220 °C (bottom) on Ag(111) and Ag(100) surfaces.

**Table S1:** Peak-fitting parameters of the O 1s peak for BTB deposited on **Ag(111) substrate.** Peak binding energy (BE) and FWHM of the Gaussian part are given for the as-deposited BTB layer (protonated), BTB annealed at 175 °C, and annealed at 235 °C (fully deprotonated). Voigt function was used for fitting; the width of the Lorentzian component was 0.1 eV.

|  | Components O1 | | Components O2 | | Component O3 (carboxylate oxygen) | |
|---|---|---|---|---|---|---|
|  | BE (eV) | FWHM (eV) | BE (eV) | FWHM (eV) | BE (eV) | FWHM (eV) |
| **Deposited** | 533.84 + 532.49 | 1.6 | 531.94 + 531.09 | 1.6 | – | – |
| **Intermediate** | 533.40 + 532.05 | 1.6 | 531.57 + 530.72 | 1.1 | 530.50 | 1.2 |
| **Deprotonated** |  |  |  |  | 530.33 | 1.1 |

Positions of Ag $3d_{5/2}$ peak were in the interval of $(368.23 \pm 0.01)$ eV for all the measurements.

**Table S2:** Peak-fitting parameters of the O 1s peak for BTB deposited on **Ag(100) substrate.** Peak binding energy (BE) and FWHM of the Gaussian part are given for the as-deposited BTB layer (protonated), BTB annealed at 120 °C, and annealed at 220 °C (fully deprotonated). Voigt function was used for fitting; the width of the Lorentzian component was 0.1 eV.

|  | Components O1 | | Components O2 | | Component O3 (carboxylate oxygen) | |
|---|---|---|---|---|---|---|
|  | BE (eV) | FWHM (eV) | BE (eV) | FWHM (eV) | BE (eV) | FWHM (eV) |
| **Deposited** | 533.78 + 532.45 | 1.6 | 531.98 + 531.13 | 1.1 | 530.55 | 1.3 |
| **Intermediate** | 533.42 + 532.07 | 1.3 | 531.64 + 530.79 | 1.1 | 530.55 | 1.1 |
| **Deprotonated** |  |  |  |  | 530.51 | 1.1 |

Positions of Ag $3d_{5/2}$ peak were in the interval of $(368.20 \pm 0.01)$ eV for all the measurements.

## 2. STM images of the as-deposited phase and decarboxylated polymeric network

Figure S3 shows the STM image of the as-deposited BTB phase on Ag(111), in which BTB molecules form a compressed ribbon-like structure. A polymeric network of decarboxylated molecules formed after sample annealing at 250 °C is shown in Figure S4.

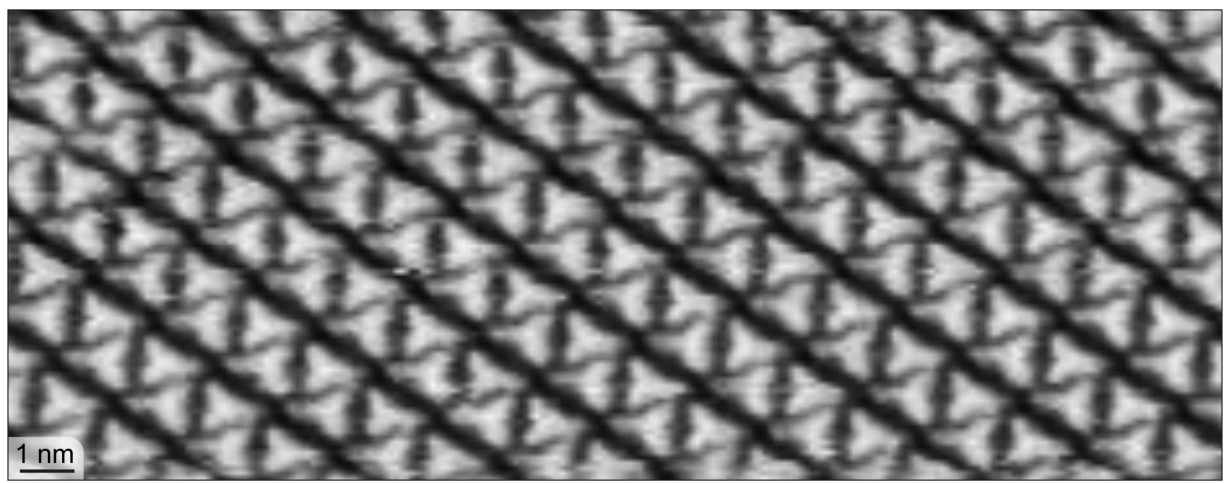

**Figure S3:** As-deposited BTB phase on Ag(111). Scanning parameters: 1.2 V, 50 pA.

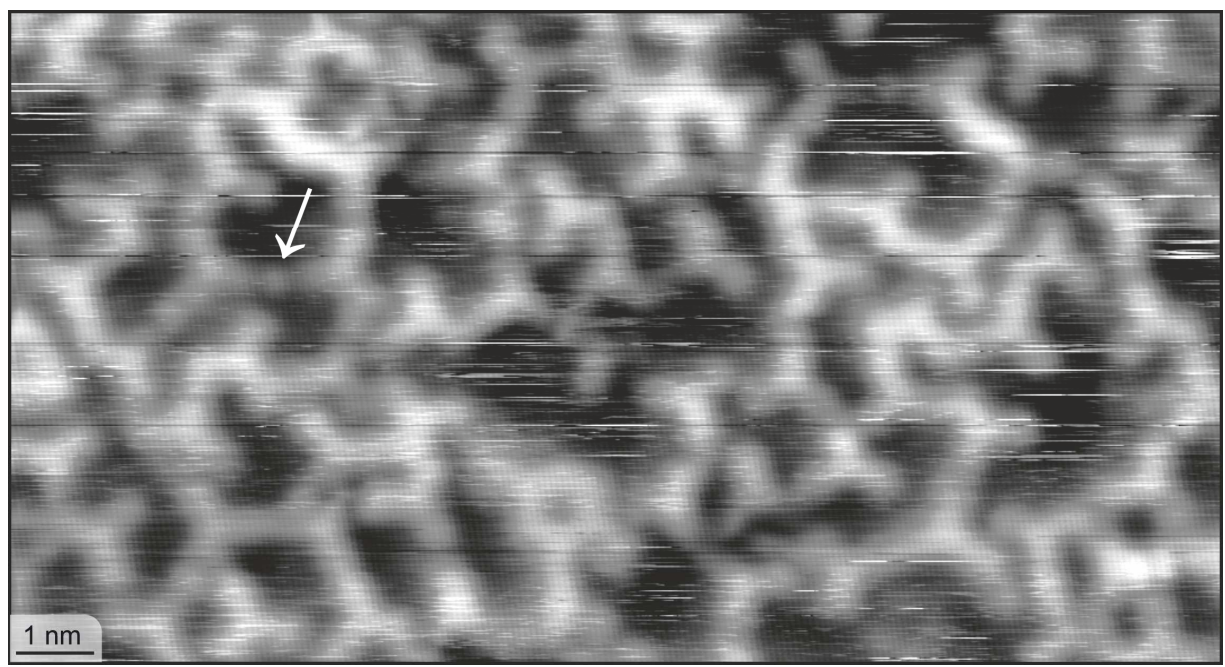

**Figure S4:** Polymeric network comprising decarboxylated BTB molecules. The circular object (highlighted by arrow) is an Ag adatom bound to phenyl radicals. Scanning parameters: 0.8 V, 100 pA.

## 3. Compact δ-BTB layer and pentacene-BTB mixed phases on Ag(100)

The robustness of the compact δ-BTB layer was also tested on Ag(100) to show the large insensitivity of the BTB layer to a surface termination. Figure S5 shows submonolayer coverage of δ-BTB on Ag(100) with a close-packed structure similar to Ag(111). The deprotonation was obtained by annealing the sample with as-deposited BTB to 180–200 °C. To obtain the full δ-BTB layer, the initial coverage was increased accordingly. The large-scale STM image of the compact δ-BTB layer on Ag(100) is given in Figure S6.

Next, by the deposition of 0.5 ML of pentacene and 0.5 ML of BTB molecules on Ag(100) and subsequent annealing, the pentacene–BTB mixed phase with a corresponding 1:1 ratio was formed. The STM and LEEM results are summarized in Figure S7. The measured diffraction and its model made in ProLEED Studio are shown in Figure S7b and c, respectively. Figure S7d provides an STM image of the mixed phase with a marked unit cell. The modeled structure and the superlattice arrangement in the matrix notation with respect to the substrate lattice are depicted in Figure S7e and f, respectively.

Similar to Ag(111), the formation of distinct mixed phases on Ag(100) is observed for different initial ratios of deposited molecules. In a separate experiment, the annealing of a mixture with different initial pentacene:BTB ratio led to the formation of two distinct molecular phases at the same time. The first is the 1:1 phase described above, and the second is 3:4. The STM and LEEM analysis is shown in Figure S8. Both molecular phases are clearly visible in the bright-field image (Figure S8a). The diffraction in Figure S8b is therefore composed of both phases, which are distinguished in the model (Figure S8c) by red (3:4 phase) and black (1:1 phase) spots. The STM image of the 3:4 phase with marked unit cell and its model are shown in Figure S8d and Figure S8e, respectively. Figure S8f shows the superlattice model of the 3:4 phase in the matrix notation.

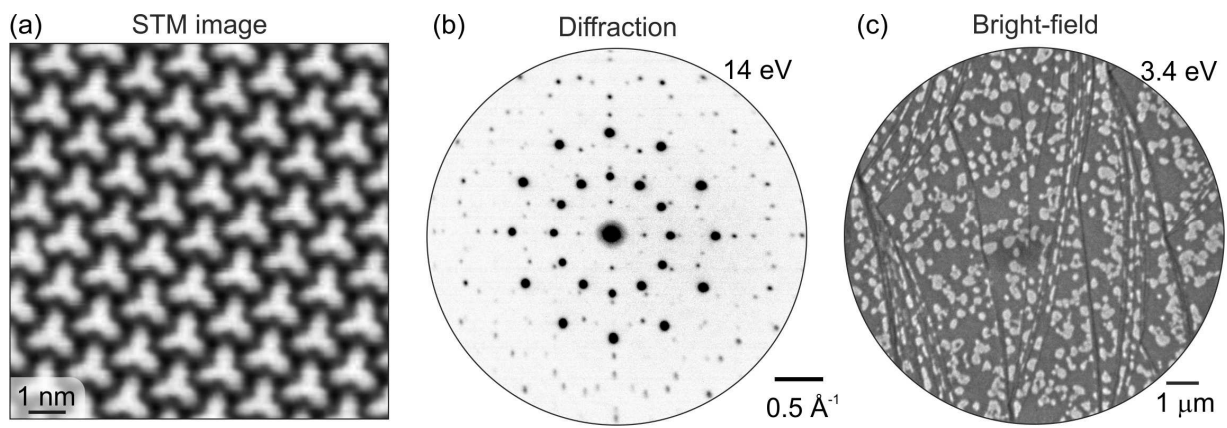

**Figure S5:** Submonolayer coverage of δ-BTB on Ag(100). (a) STM, (b) diffraction, and (c) LEEM bright-field image. Scanning parameters in (a): 1.0 V, 50 pA.

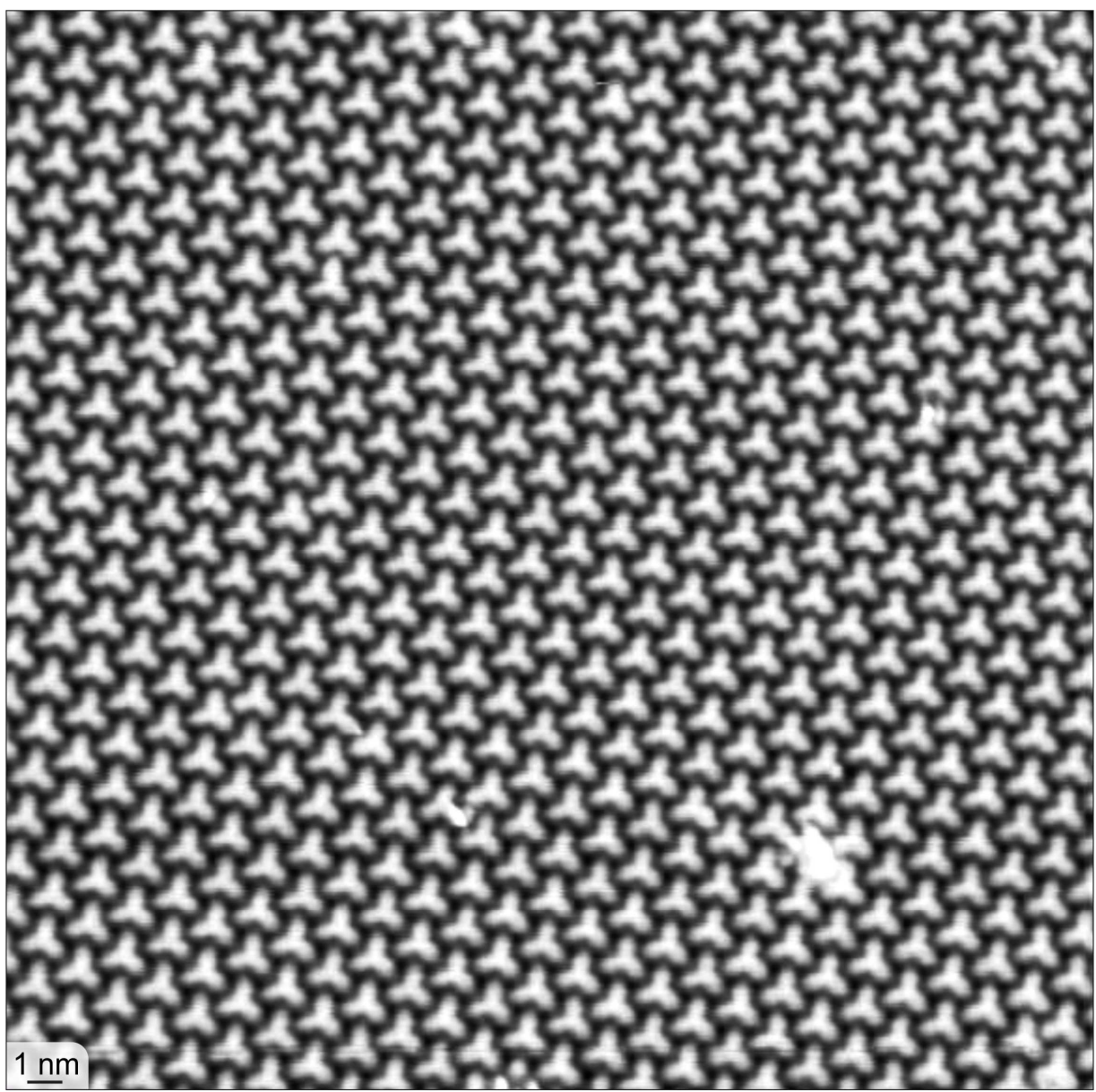

**Figure S6:** Compact δ-BTB layer on Ag(100) surface. Scanning parameters: 1.0 V, 50 pA.

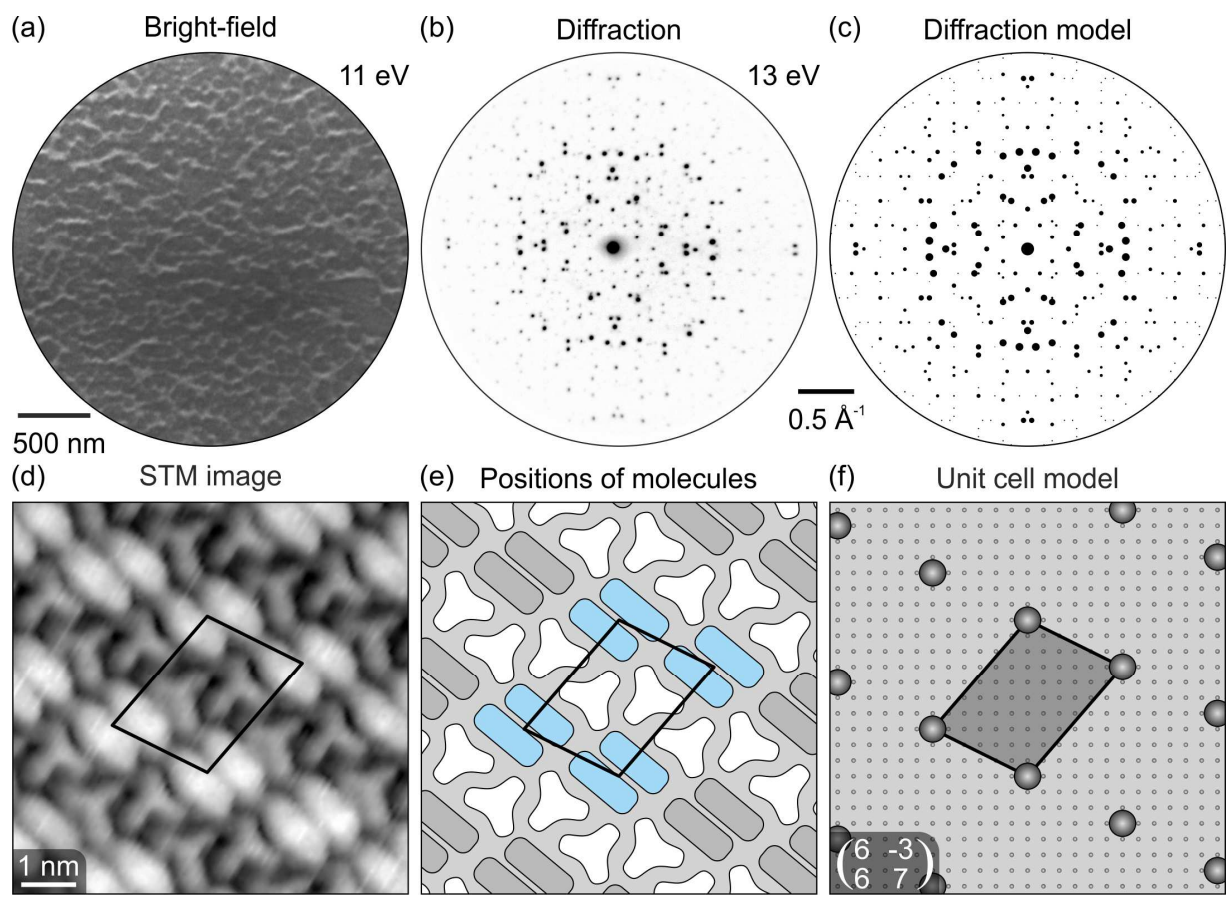

**Figure S7:** STM and LEEM analysis of pentacene-BTB mixed phase with 1:1 ratio of pentacene and BTB at submonolayer coverage. (a) Bright-field LEEM, (b) diffraction, and (c) the modeled diffraction. (d) STM image of the 1:1 structure, (e) modeled positions of the molecules, and (f) their corresponding superlattice description in the matrix notation.

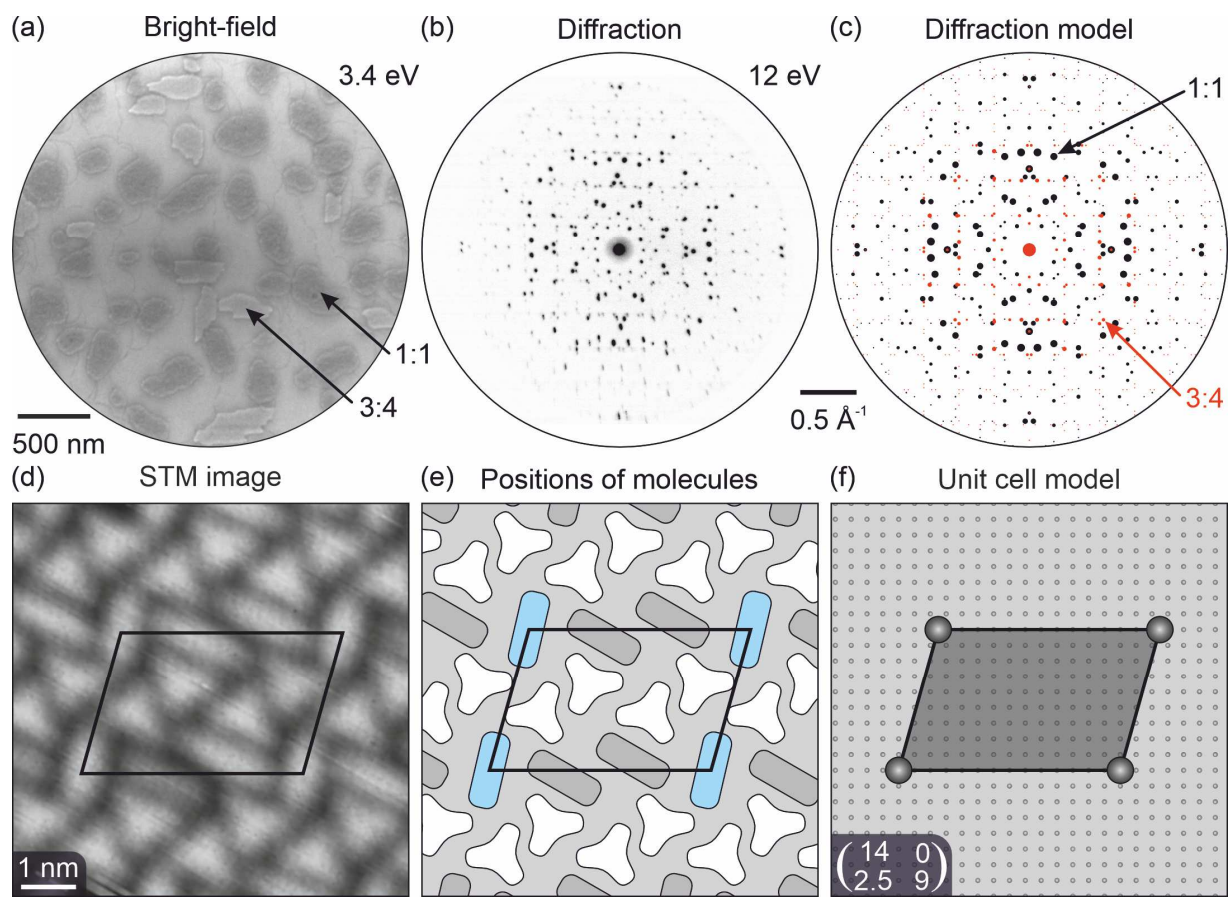

**Figure S8:** STM and LEEM analysis of pentacene-BTB mixed phase with 3:4 pentacene to BTB ratio at submonolayer coverage. (a) Bright-field LEEM, (b) diffraction, and (c) the modeled diffraction shows the combination of 3:4 and 1:1 mixed phases both present on the surface. The 3:4 phase diffraction model is red in (c). (d) STM image of the 3:4 structure, (e) modeled positions of the molecules, and (f) their corresponding superlattice description in the matrix notation.

## 4. Laboratory XPS analysis of the compact δ-BTB layer

The O 1s and C 1s photoemission spectra of the fully deprotonated compact δ-BTB layer on Ag(111) substrate measured in situ by a laboratory XPS (Figure S9) are consistent with the synchrotron measurements. The photoelectron spectrum for the O 1s region shows a single peak at binding energy (BE) of 530.5 eV, consistent with the fully deprotonated BTB on the surface. The C 1s region comprises a peak component associated with phenyl rings at BE 284.2 eV and a carboxyl peak at BE ~287.5 eV, again fully consistent with synchrotron radiation spectra.

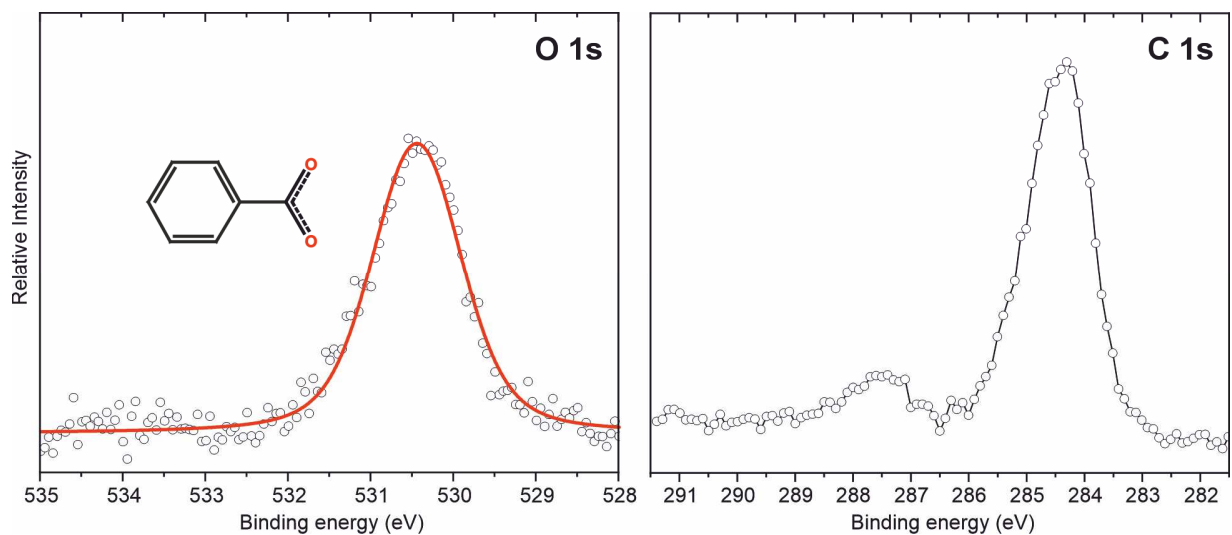

**Figure S9:** O 1s (left) and C 1s (right) measured by laboratory XPS on compact δ-BTB layer on Ag(111).

## 5. Supplementary STM figures of δ-BTB on Ag(111)

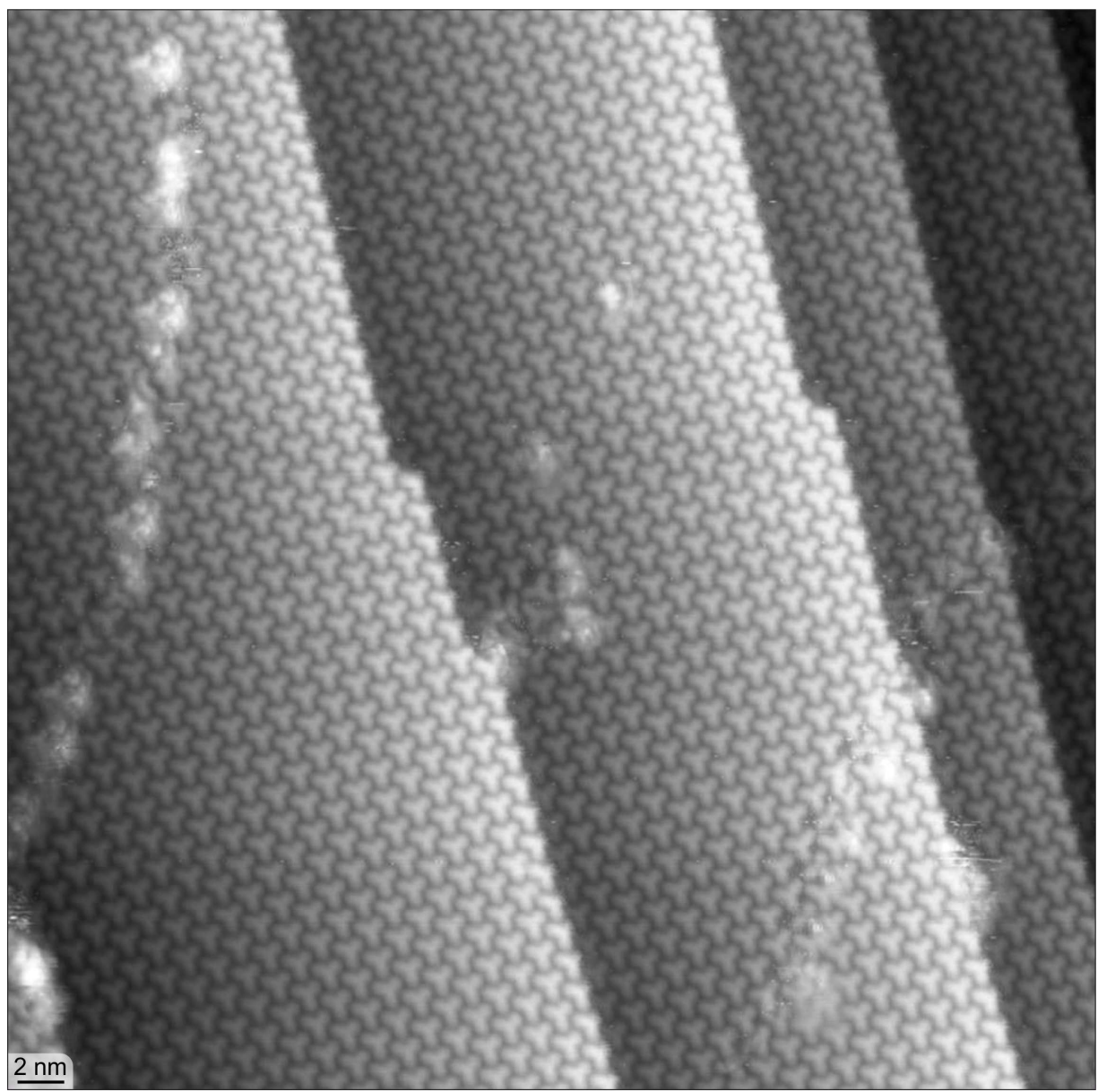

**Figure S10:** Large-scale STM image of compact δ-BTB layer on Ag(111) seamlessly extending over several terraces. Scanning parameters: 1.4 V, 50 pA.

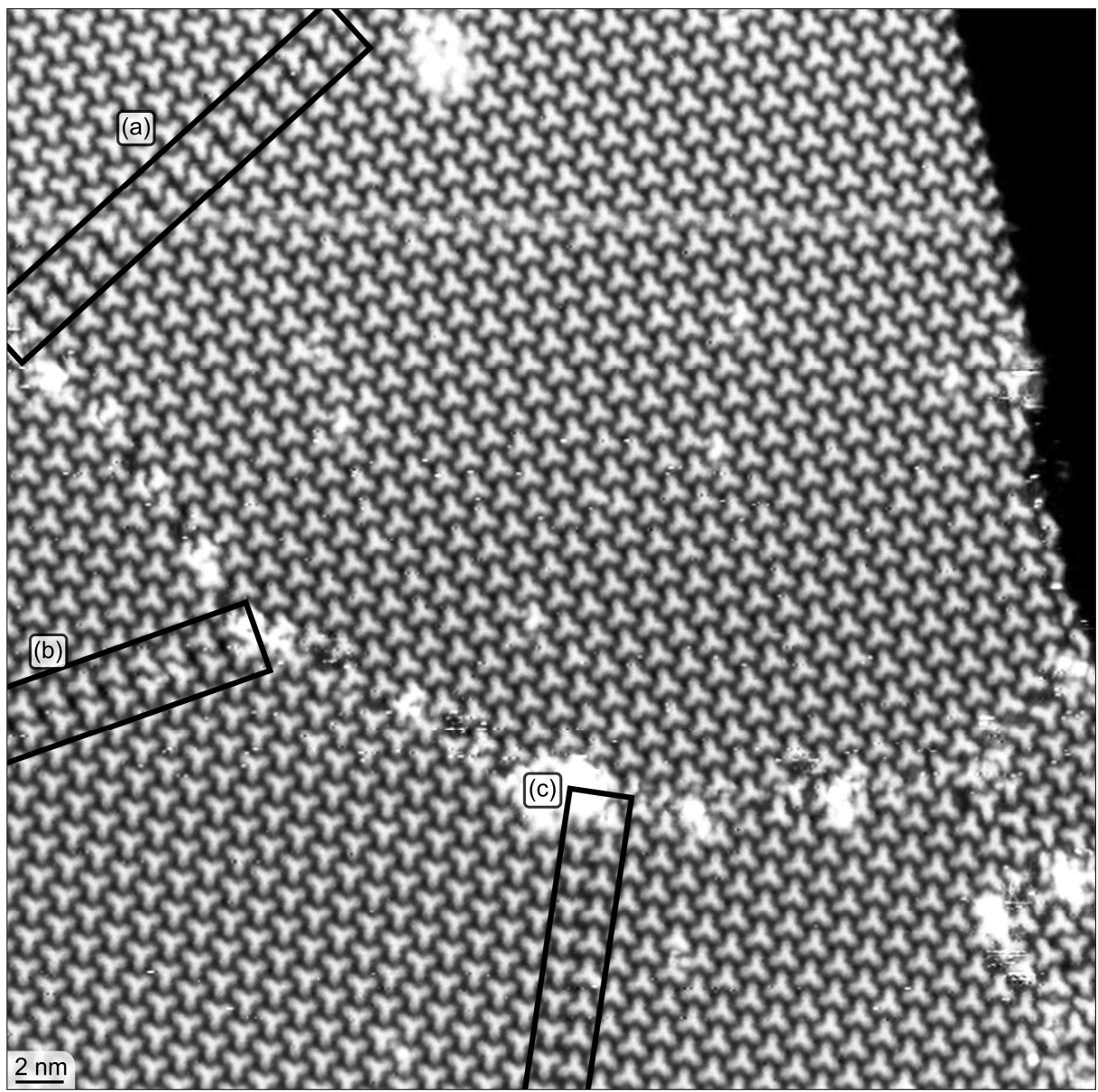

**Figure S11:** STM image of compact δ-BTB layer on Ag(111). Domain boundaries are marked with black rectangles (a–c). Scanning parameters: 1.4 V, 50 pA.

## 6. Comparison of δ-BTB unit cell on Ag(100) and Ag(111)

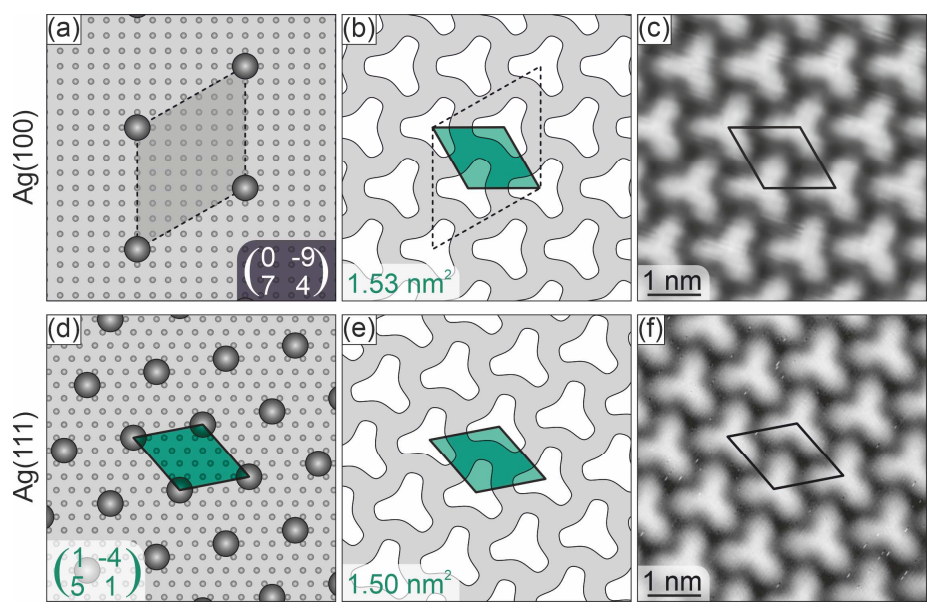

**Figure S12:** Comparison of the structure of the compact δ-BTB layer for (a–c) Ag(100) and (d–f) Ag(111). (a) The alignment of the commensurate δ-BTB unit cell with the Ag(100) lattice. (b) Position of the δ-BTB molecules within the unit cell. (c) The STM image of the δ-BTB layer on Ag(100) with highlighted apparent unit cell given in green in panel (b). (d) The commensurate unit cell (green) for δ-BTB on Ag(111). (e) Position of the δ-BTB molecules within the unit cell. (f) STM image confirming the presented model.

## 7. Stable overlayers of HM-TP and HAT-CN

In addition to pentacene, we have also tested two other organic semiconductors, i.e., 2,3,6,7,10,11-hexamethoxytriphenylene (HM-TP) and 1,4,5,8,9,12-hexaazatriphenylene hexacarbonitrile (HAT-CN), to show the generality of the robustness of the compact δ-BTB layer with respect to mixing with deposited overlayers.

Synchrotron radiation photoelectron spectroscopy measurements on a compact δ-BTB layer on Ag(111) acquired before deposition, after deposition of submonolayer coverage and subsequent annealing at 220 °C are given in Figure S13 for HAT-CN and Figure S14 for HM-TP. In the case of HAT-CN, we observe a new peak at 287.2 eV after HAT-CN deposition; after annealing, its intensity decreases, but HAT-CN is not removed completely below the decarboxylation threshold of BTB. As HAT-CN does not contain oxygen atoms, we observe a decrease in the O 1s signal after HAT-CN deposition due to covering the δ-BTB layer and attenuation of the signal from the lower-lying layers. After annealing to 220 °C, the O 1s peak intensity slightly increases. Importantly, we do not observe any significant changes in both peak position and shape that would indicate changes in the δ-BTB layer, i.e., its interaction with HAT-CN.

In the case of HM-TP, our photoelectron spectroscopy measurements show a similar result. The spectra in Figure S14 show that after HM-TP deposition, there is an additional signal in the C 1s spectrum and an additional single peak at 533.9 eV in the O 1s spectrum. The peak associated with carboxylate oxygen is only slightly reduced in intensity, and we do not observe any changes in its position or shape. After annealing at 220 °C, both C 1s and O 1s spectra almost change back to their original appearance; however, a small portion of HM-TP molecules remain on the surface. Also, in this case, we do not observe any signs of change in peaks associated with the δ-BTB layer.

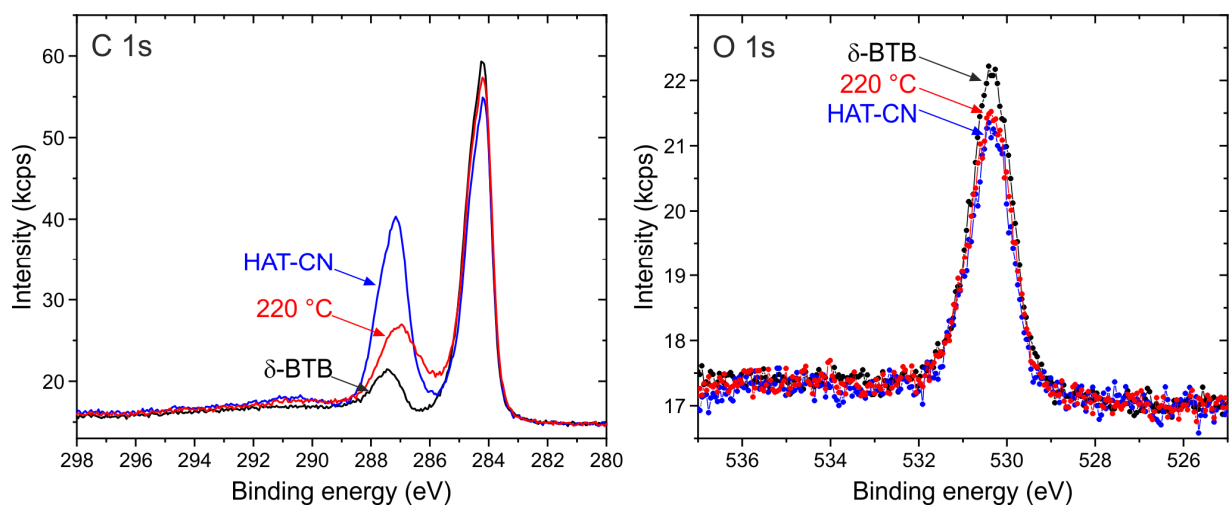

**Figure S13:** C 1s and O 1s spectra obtained by synchrotron radiation photoelectron spectroscopy for HAT-CN deposited on compact δ-BTB layer on Ag(111). The spectra measured on a compact δ-BTB layer are marked as "δ-BTB", spectra measured after a deposition "HAT-CN" and after subsequent annealing to 220 °C "220 °C". The spectra are vertically shifted to match the background at the low binding energy side of each peak.

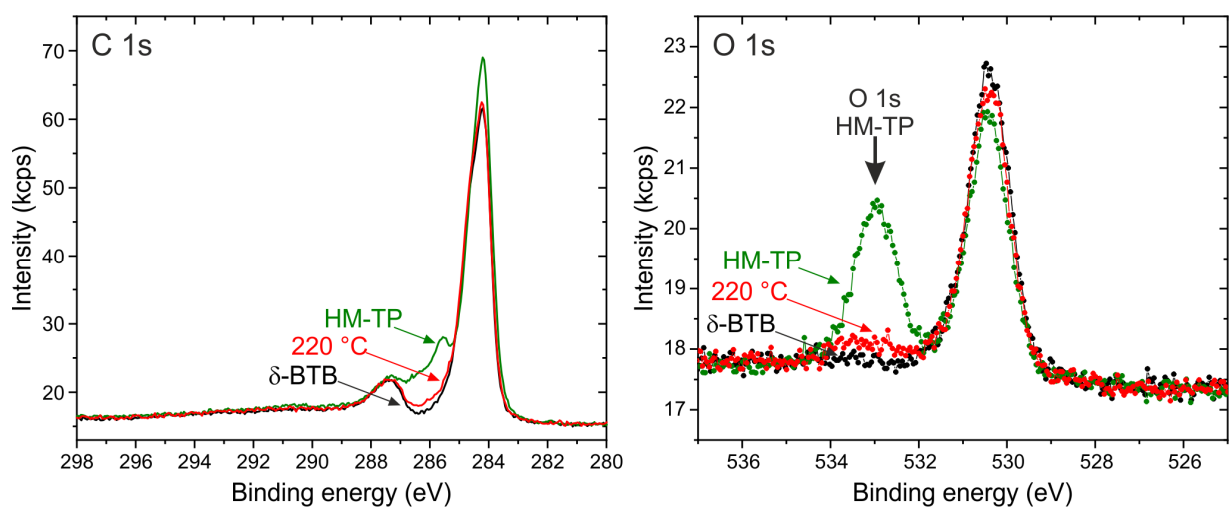

**Figure S14:** C 1s and O 1s spectra obtained by synchrotron radiation photoelectron spectroscopy for HM-TP deposited on compact δ-BTB layer on Ag(111). The spectra measured on a compact δ-BTB layer are marked as "δ-BTB", spectra measured after a deposition "HM-TP" and after subsequent annealing to 220 °C "220 °C". The spectra are vertically shifted to match the background at the low binding energy side of each peak. A vertical arrow marks the O 1s peak associated with HM-TP.

LEEM analysis of the δ-BTB layer on the Ag(111) surface before and after HAT-CN deposition is summarized in Figure S15. After the submonolayer deposition, large dark areas with different microdiffraction patterns appeared in the bright-field images (Figure S15b). Therefore, we associate these areas with the ordered HAT-CN overlayer. In line with the XPS results, annealing of the substrate up to 220 °C led to a decreased area covered by the HAT-CN islands. During the experiment, we did not observe any signs of disruption of the compact δ-BTB layer or formation of mixed HAT-CN–BTB phases. However, as HAT-CN remains on the surface, an unambiguous proof of not mixing cannot be provided solely by LEEM measurements.

Similarly, we have tested HM-TP, but in this case, the LEEM analysis is complicated by the fact that HM-PT has the same unit cell as the δ-BTB layer. Also, in this case, we did not observe any signs of disruption of the compact δ-BTB layer.

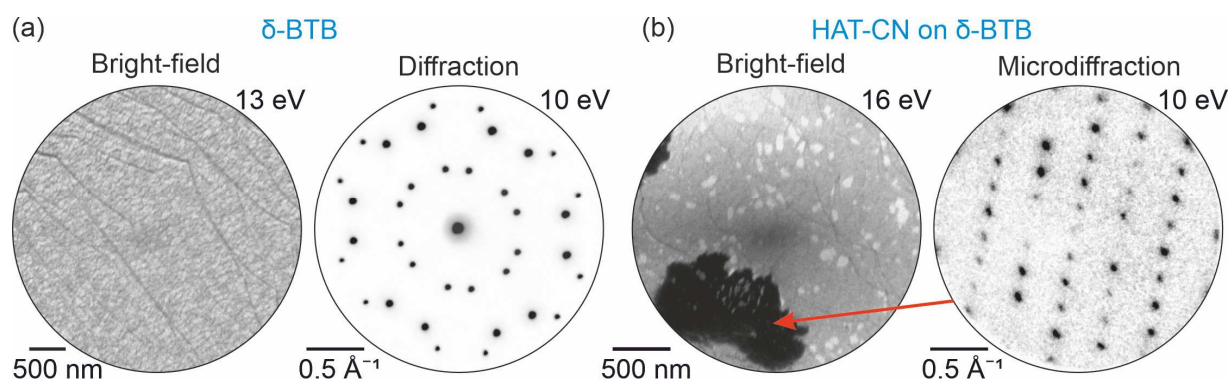

**Figure S15:** LEEM bright-field and diffraction pattern images before (a) and after (b) HAT-CN deposition. Microdiffraction images taken from the HAT-CN overlayer indicate different periodicity of HAT-CN islands.

## 8. The wheel-like pentacene-BTB mixed phase on Ag(111)

In addition to the 1:1 pentacene-BTB mixed phase discussed in the main text, 2:1 phase was also obtained for a different initial pentacene:BTB ratio. The bright-field LEEM given in Figure S16a shows bright islands of the mixed phase on Ag(111) substrate. Figures S16b and c show the mixed phase diffraction and its model. The size and orientation of the unit cell fit the real space structure observed in the STM image of the mixed phase given in Figure S16d. The unit cell consists of 6 pentacene and 6 δ-BTB molecules arranged into a wheel, and 6 additional pentacenes at its periphery (see Figure S16e). The position of the unit cell with respect to the substrate lattice is given in Figure S16f.

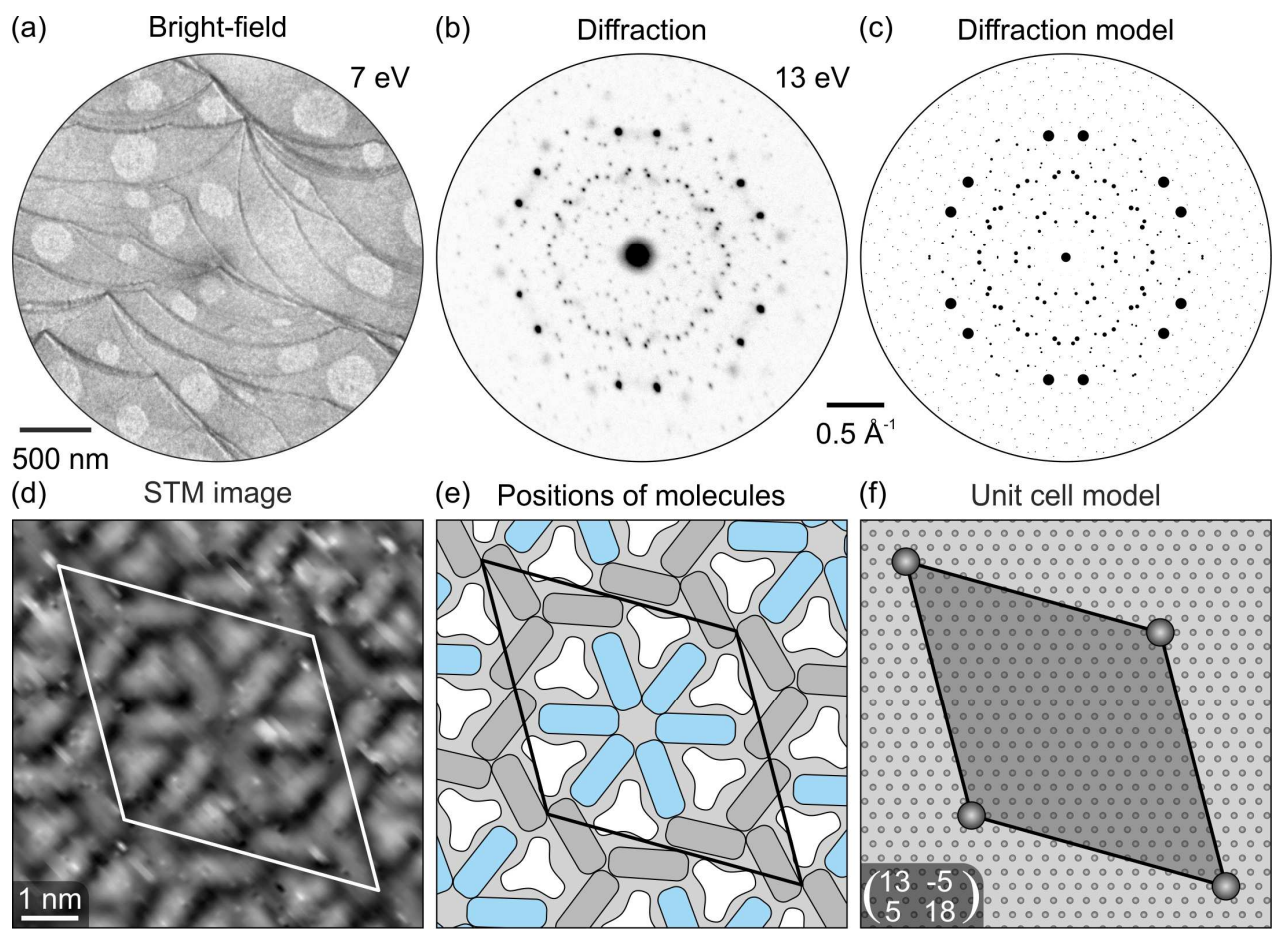

**Figure S16:** The wheel-like pentacene-BTB mixed phase on Ag(111) with 2:1 ratio of pentacene:BTB. (a) bright field LEEM, (b) diffraction pattern, and (c) modeled diffraction of the mixed phase. (d) STM image with unit cell marked by a black rhombus. Scanning parameters: 1.1 V, 50 pA. (e) Schematics of positions of the molecules within a unit cell. (f) Superstructure unit cell with respect to the substrate.

## 9. Gas-phase reference for calculated adsorption energies

The stability of the δ-BTB layer is evaluated in the main text with respect to direct desorption of an ionized BTB molecule. An alternative desorption process of a BTB molecule from the δ-BTB layer consists of a full protonation of a BTB molecule, followed by its detachment from the molecular layer into a vacuum. Here, a fully protonated gas-phase BTB molecule would serve as the reference state. In the following we demonstrate that the δ-BTB phase is favored over the intermixed phase even in this case, although the quantitative outcomes are strongly functional dependent. In this scenario, the adsorption energy per unit area $\gamma$ is a function of a hydrogen chemical potential:

$$\gamma\left(\mu_{H_2}\right) = \frac{E_{int} - E_{sub} + \frac{3}{2}\left(E_{H_2} + \mu_{H_2}\right) - E_{H_3BTB}^{gas}}{S}, \qquad (S1)$$

where $E_{int}$ and $E_{sub}$ are total energies of an interface and a Ag(111) substrate system, $E_{H_2}$ and $E_{H_3BTB}^{gas}$ are gas phase energies for a hydrogen molecule and for a (fully protonated) BTB molecule, $\mu_{H_2}$ is a hydrogen molecule chemical potential, and $S$ stands for an area of the supercell.

Calculated adsorption energies per unit area employing PBE-D3 and optB86b functionals are depicted in Figure S17. As illustrated, under experimental conditions related to hydrogen chemical potential of –1.07 eV, the δ-BTB phase is calculated to be in both cases more stable than the intermixed phase. Nonetheless, the calculated critical chemical potentials and transition pressures exhibit significant dependence on the DFT functional employed. The transition occurs at –0.89 eV (PBE-D3) and at –0.49 eV (optB86b), which corresponds to the hydrogen transition pressure of ~$10^{-7}$ mbar and ~1 mbar, respectively. We also note that the phase diagram is not complete, as it does not contain semi-protonated and protonated BTB phases. However, given the seven orders of magnitude difference in the calculated critical pressure yielded by these functionals, we conclude that DFT does not provide a deeper understanding of the phase stability as a function of the hydrogen chemical potential.

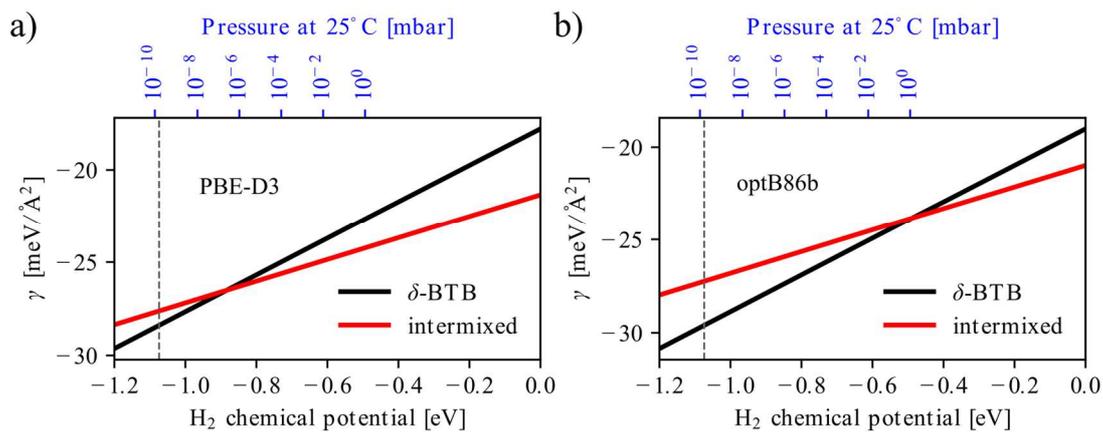

**Figure S17:** Calculated surface energies of the δ-BTB and intermixed phases as a function of a hydrogen chemical potential and a hydrogen pressure at 25 ˚C using (a) PBE-D3 functional and (b) optB86b functional. A gray dashed line marks a chemical potential under experimental conditions.

## 10. Supplementary DFT structures of molecular monolayers on Ag(111) substrate

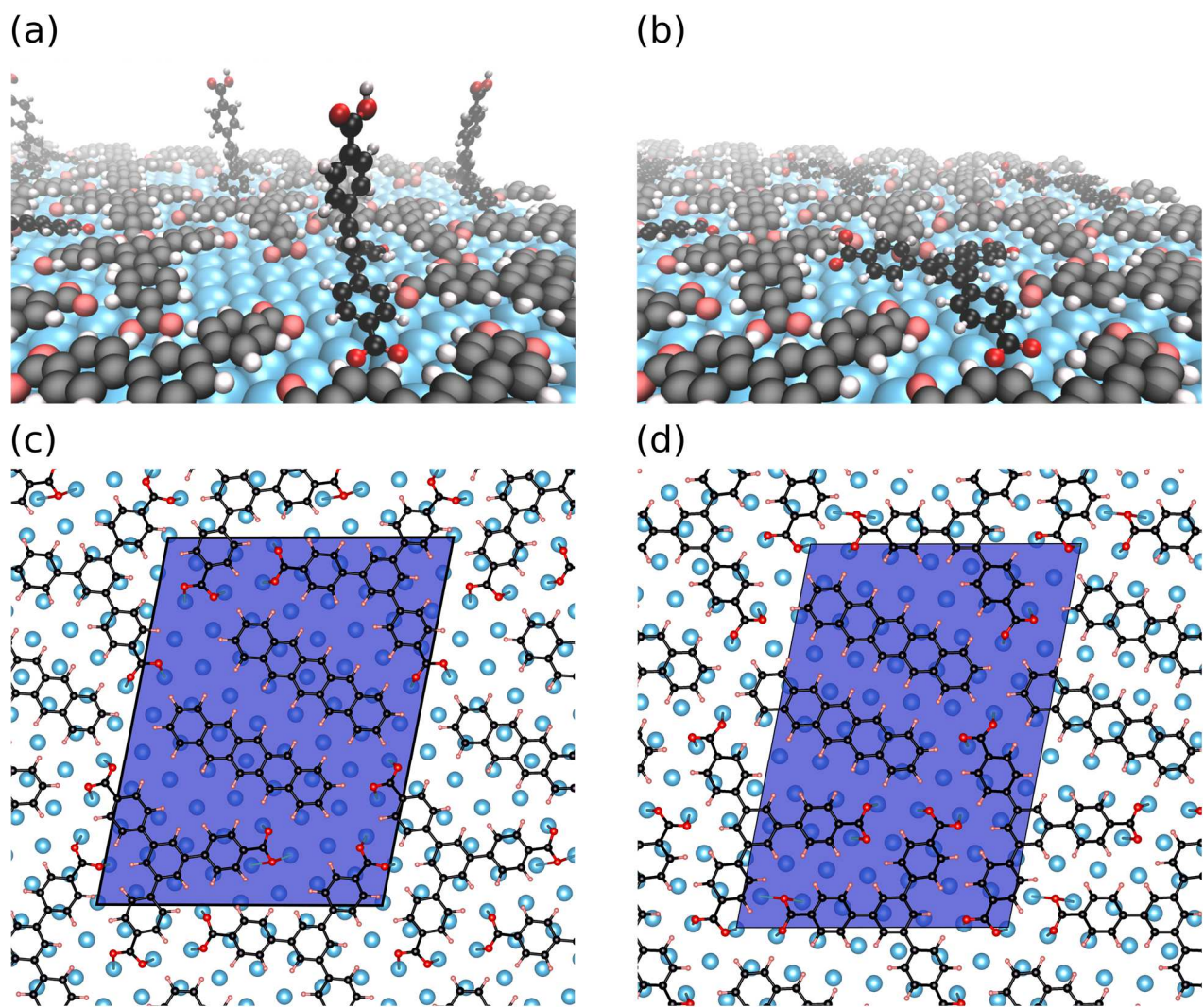

**Figure S18:** Models used for DFT calculations. (a) and (b) partially protonated BTB molecule in the standing-up (a) and in the lying (b) configurations. The BTB molecule being detached is rendered with higher contrast. (c) Structural model for the intermixed phase with the unit cell obtained from diffraction. (d) The same structure with the unit cell giving the highest stability.